\documentclass[12pt]{article}

\usepackage[fleqn]{amsmath}
\usepackage{amsfonts}
\usepackage{amssymb}
\usepackage{epsfig}
\usepackage[a4paper]{geometry}
\catcode233=13   \def é{\'e}
\catcode232=13   \def è{\`e}
\catcode234=13   \def ê{\^e}
\catcode235=13   \def ë{\"e}

\catcode225=13   \def á{\'a}
\catcode224=13   \def à{\`a}
\catcode226=13   \def â{\^a}
\catcode228=13   \def ä{\"a}

\catcode239=13   \def ï{\"{\i}}
\catcode238=13   \def î{\^{\i}}
\catcode244=13   \def ô{\^{o}}
\catcode252=13   \def ü{\"{u}}
\catcode249=13   \def ù{\`{u}}

\catcode192=13   \def À{\`{A}}
\catcode201=13   \def É{\'{E}} 
\catcode200=13   \def È{\`{E}}
\catcode217=13   \def Ù{\`{u}}
\catcode231=13   \def ç{\c{c}}

\newcommand{\vect}[1]{\boldsymbol{#1}} %{\vec{#1}}    %{\bbox{#1}}        %
\newcommand{\TF}[1]{\widetilde{#1}}

\newlength{\largeur}

\newtheorem{lemma}{Lemma}

\newcounter{preuve}

\newcommand{\CQFD}{$\Box$}

\newcommand{\al}{\alpha}
\newcommand{\be}{\beta}

\newcommand{\eps}{\epsilon}
\newcommand{\sig}{\sigma}
\newcommand{\lam}{\lambda}

\newcommand{\chia}{\chi_{a}}
\newcommand{\chib}{\chi_{b}}
\newcommand{\taua}{\tau_{a}}
\newcommand{\taub}{\tau_{b}}

\newcommand{\lamD}{\lambda_{\text{\mbox{\fontsize{6}{6}\selectfont D}}}}

\newcommand{\rmH}{_{\text{\mbox{\fontsize{6}{6}\selectfont H}}}}

\newcommand{\IR}{{\mathbb R}}

\newcommand{\OO}{{\cal O}}

\newcommand{\CC}{{\cal C}}
\newcommand{\LL}{{\cal L}}

\newcommand{\FF}{{\cal F}}

\newcommand{\vz}{\vect{0}}
\newcommand{\ver}{\vect{r}}
\newcommand{\vx}{\vect{x}}
\newcommand{\vk}{\vect{k}}

\newcommand{\vp}{\vect{p}}

\newcommand{\vP}{\vect{P}}

\newcommand{\noi}{\noindent}

\newcommand{\rom}[1]{\mathrm{#1}}

\newcommand{\grad}{\vect{\nabla}}

\newcommand{\eq}[1]{\begin{equation} #1 \end{equation}}

\newcommand{\e}[1]{{\mathrm e}^{#1}}

\newcommand{\Tr}{\rom{Tr}}
\newcommand{\tend}{\rightarrow}

\newcommand{\bracket}[1]{\left\langle #1 \right\rangle}
\newcommand{\acc}[1]{\left\{ #1 \right\}}
\newcommand{\parenth}[1]{\left( #1 \right)}

\newcommand{\absol}[1]{\left| #1 \right|}

\newcommand{\romd}{\rom{d}}
\newcommand{\romD}{\rom{D}}

\newcommand{\ide}{^{\text{id}}}

\newcommand{\TT}{\text{\mbox{\fontsize{6}{6}\selectfont T}}}

\newcommand{\el}{_{\text{e}}}
\newcommand{\pr}{_{\text{p}}}

\newcommand{\R}{\vect{R}}

\newcommand{\X}{\vect{X}}
\newcommand{\vxi}{\vect{\xi}}

\newcommand{\ket}[1]{\left| #1 \right\rangle}
\newcommand{\bra}[1]{\left\langle #1 \right|}

\newcommand{\FC}{F_{\rm c}}

\newcommand{\Graph}[1]{ \raisebox{-0.49 \height}{\epsfig{file=#1}} }

\newcommand{\PreserveBackslash}[1]{\let\temp=\\#1\let\\=\temp}

\newcommand{\chiTk}{\TF{\chi}(\vk)}

\newcommand{\phich}{\phi_{\text{ch}}}

\newcommand{\op}[1]{\rom{#1}}
\newcommand{\id}{1\!{\mathrm l}}
\newcommand{\Vo}{\op{V}_{\mbox{\fontsize{7}{7}\selectfont(0)}}}

\newcommand{\Ao}{\op{A}_{\mbox{\fontsize{7}{7}\selectfont(0)}}}
\newcommand{\opq}{{\text{{\bf q}}}}
\newcommand{\opr}{{\text{{\bf r}}}}
\newcommand{\opp}{{\text{{\bf p}}}}
\newcommand{\opR}{{\text{{\bf R}}}}
\newcommand{\opU}{\op{U}}
\newcommand{\opA}{\op{A}}
\newcommand{\opB}{\op{B}}

\newcommand{\opP}{\op{P}}
\newcommand{\opQ}{\op{Q}}

\newcommand{\opV}{\op{V}}

\newcommand{\opH}{\op{H}}

\newcommand{\opT}{{\cal T}}

\newcommand{\adj}{^{\dagger}}
\newcommand{\ignore}[1]{}

\newcommand{\n}{{ n}}		%\sf
\newcommand{\ab}{a_{\text{\mbox{\fontsize{6}{6}\selectfont B}}}}
\newcommand{\aB}{a_{\text{\mbox{\fontsize{6}{6}\selectfont B}}}}

\newcommand{\Reste}{^{\text{(D)}}}

\newcommand{\FR}{F_{\rm R}}

\newcommand{\chiTMFk}{\TF{\chi}_{\rm MF}(\vk)}

\usepackage{enumerate}
\usepackage{psfrag}
\usepackage{mathrsfs}     %super ecriture calligraphique

\renewcommand{\n}{\mathsf{n}}
\newcommand{\m}{\mathsf{m}}

\newcommand{\atH}{_{\text{a}}}
\newcommand{\Ea}{E\atH}
\newcommand{\rhoat}{\rho\atH^{\text{id}}}
\newcommand{\chiTatk}{\TF{\chi}_{\text{at}}(\vk)}
\newcommand{\BB}{{\cal B}}

\newcommand{\lamat}{\lam\atH}
\newcommand{\rmp}{{\rm p}}
\newcommand{\rme}{{\rm e}}
\newcommand{\alphaH}{\alpha\rmH}

\begin{document}

\title{Dielectric versus conductive behaviour in quantum gases: exact results for the hydrogen plasma}
\author{
V. Ballenegger\thanks{Corresponding author. Present~address: Department of Chemistry, University of Cambridge, Lensfield Road, Cambridge CB2 1EW, United Kingdom. E-mail: {\tt vcb25@cam.ac.uk}}\ \thanks{Supported by the Swiss National Foundation for Scientific Research.}\ \ and Ph. A. Martin\\
\footnotesize \it Institut de th\'eorie des ph\'enom\`enes physiques, EPFL\\
\footnotesize \it \'Ecole Polytechnique F\'ed\'erale de Lausanne\\
\footnotesize \it CH-1015 Lausanne, \textsc{Switzerland}.
}

\maketitle

\abstract{We study the electrical susceptibility of a hydrogen gas at equilibrium, partially ionized by thermal excitations. The gas is described as a quantum plasma of point protons and electrons, interacting via the Coulomb potential. Using the newly developped diagrammatical technique of screened cluster expansions, we calculate exactly the wavenumber-dependent susceptibility in the atomic limit, where most charges are bound into hydrogen atoms. A transition from conductive to dielectric behaviour occurs when the wave length is decreased well below the Debye screening length. The standard formula for the dielectric function of an ideal gas of hydrogen atoms is recovered in an appropriate scaling limit. The derivation treats all effects arising from the Coulomb interaction (screening, binding, polarization) in a fully coherent way, without intermediate approximation nor modelization.
}\\

\noi PACS:
  52.25.Mq; \ 	%Dielectric properties (Physics of plasmas and electric discharges)
  52.25.Jm; \ 	%Ionization of plasmas
%  05.30.-d	Quantum statistical mechanics


\section{Introduction}

Screening in a non-relativistic nucleo-electronic plasma is one of the most 
important consequence of the long range of the Coulomb force. For quantum
mechanical point charges, there is a number of physical effects that can
participate to the screening mechanisms. First, there is a collective screening
phenomenon which leads to the formation of neutralizing polarization clouds (as in the classical plasma). The clouds are made of unbound (ionized) charges, and are
always present at any non zero temperature. The spatial extension of such clouds
is of the order, say, of the Debye-H\"{u}ckel screening length. On the scale of the Bohr radius, chemical binding may lead to the formation of neutral atoms or molecules.
Finally, in a partially recombined plasma, there exists also a dielectric screening due to the
polarization of atomic dipoles and molecules. 

In usual theories, these different phenomena
lead to different ways of modelling the system, each of them having its own
range of validity. In a conducting phase, the ionic screening due to the unbound charges
may be taken into account by replacing in calculations the bare Coulomb
potential by a mean field potential obtained within the Debye-H\"uckel or
Random Phase Approximation. If one is interested in a dielectric phase, 
it is often appropriate to neglect ionic screening and to separate the problem
into two parts. The role of quantum mechanics is limited to an a priori
calculation of the polarizability of a single atom, and then the problem is
treated in the framework of classical statistical mechanics of preformed dipoles,
characterized by these quantum mechanical atomic data. It is nevertheless true
that all these effects have a single common origin, the Coulomb interaction.  
All of them should stem in a consistent way from the basic  $N$-body Hamiltonian
\eq{
\opH=\sum_{j=1}^{N}\frac{|{\opp}_{j}|^{2}}{2m_{\alpha_{j}}}+\sum_{i<j}^{N}
e_{\alpha_{i}}e_{\alpha_{j}} V(\opr_{i}-\opr_{j}),
\qquad V(\opr) = \frac{1}{|\opr|},
\label{A2 0}
}
describing the Coulomb interaction of $N$ quantum particles (point nuclei and
electrons) of species $\alpha=1,2,\ldots$ with charges $e_{\alpha}$ and masses
$m_{\alpha}$. Such a fundamental attitude is legitimate not only as a matter of principle,
but also because of the need to a have a coherent scheme for understanding
the interplay and relative importance of these effects. 

In this paper, we present a detailed study of the response function $\chi(\bf r)$ of a quantum plasma to a classical localized external charge density $c_{\rm ext}(\bf r)$. Its Fourier transform
\eq{	\label{A2 def chi}
\chiTk=\frac{\TF{c}_{\rm ind}({\vk})}{\TF{c}_{\rm ext}({\vk})}                                  
}
is defined as the ratio, to linear order, of the induced charge $\TF{c}_{\rm ind}({\bf k})$ in the plasma to the external charge $\TF{c}_{\rm ext}({\vk})$ at wave length ${\vk}$.
It is related to the dielectric function $\epsilon({\bf k})$ by
\begin{equation} 
\chiTk=\epsilon^{-1}(\vk)-1
\label{A2 2}
\end{equation}
so that purely metallic behaviour $(\epsilon(0)=\infty)$ is characterized by the perfect screening relation
\eq{
\lim_{{k}\tend 0} \chiTk=-1
\label{A2 -1}
}
The response function can be expressed in terms of the charge fluctuations in the plasma by
\eq{
\chiTk=-\frac{4\pi}{|\vk|^{2}}\int_{0}^{\beta}\romd\tau
S(\vk,\tau).
\label{A2 4}
}
In~\eqref{A2 4}, $S(\vk,\tau)$ is the imaginary time displaced charge charge correlation function (the charge structure factor of the quantum plasma at imaginary time~$\tau$), and $\beta=(k_B T)^{-1}$ is the inverse temperature. These formulae hold in a fluid phase and it will be assumed throughout this paper that the system is spatially uniform.
The relation (\ref{A2 -1}) together with~\eqref{A2 4} is the quantum analogue of the Stillinger-Lovett perfect screening condition for a classical plasma~\cite{M}.

In a mean field treatment, the low wave number behaviour of $\chiTk$ can be represented by the Debye-H\"{u}ckel type formula
\eq{	\label{A2 D-H}
    \chiTk \simeq - \frac{\kappa^2}{k^2+\kappa^2},
}
where $\kappa^{-1}=\lamD$ is the Debye length. 
In an uniform state of the quantum system, it turns out that the relation (\ref{A2 -1}) is always true at any non-zero temperature. This result can be obtained from an analysis of the constraints imposed by the hierarchy equations for imaginary time Green's functions~\cite{MO,AM}. It can as well be derived in the formalism of charged loops (section~4) using various schemes for summing  Mayer graphs~\cite{Cornu98}. Hence, an infinitely extended quantum plasma is \emph{formally} always conducting, even in a phase composed mainly of neutral entities (atoms, molecules). This is due to the tiny amount of free charges that are present by thermal ionization.

The main question we adress in this paper is: \emph{under what conditions does the system of quantum charges exhibit a dielectric behaviour~?} To keep the discussion reasonnably simple, we restrict it to the electron-proton (e-p)
system with $N\pr$ protons, $N\el$ electrons and Hamiltonian $H_{N\pr N\el}$. 
If electrons and protons recombine into a dilute gas of hydrogen atoms with density  $\rho\rmH$, one expects, at an elementary level, the system to be characterized by a dielectric constant   
\eq{
\epsilon \simeq 1+4\pi \rho\rmH \alphaH
\label{A2 6}
}
where $\alphaH$ is the polarizability of the hydrogen atom.
In order to establish such a result starting from the many-body Hamiltonian $H_{N\pr N\el}$, it is necessary to give a precise meaning to the recombination of protons and electrons into hydrogen atoms. This is formulated in the so-called atomic limit described in section~2. In this limit one lets the temperature and the density tend to zero in a coupled way. Low temperature favors binding over ionization, whereas low density, by increasing the available phase space, favors dissociation. The rate at which the density is reduced as $T\tend 0$ determines an entropy-energy balance that selects the formation of certain chemical species. If this rate is within a certain range, hydrogen atoms are formed, and it can rigorously be shown that the equation of state becomes asymptotic to that of an ideal gas of hydrogen atoms~\cite{F1,F2,CLY}.

In order to put the issues in proper perspective, we give in section~3 a naive but mathematically ill defined derivation of (\ref{A2 6}), disregarding all collective screening effects tied with the long range of the Coulomb potential. This also enables us to specify the range of wave numbers ${\bf k}$ for which (\ref{A2 6}) is expected to hold: in view of (\ref{A2 -1}) ${\bf k}$ should not be too small to avoid the perfect screening regime, but also not too large so that the atom experiences an uniform perturbation on the scale of the extension of its center of mass wave packet (see (\ref{A2 plateau})).

The rest of the paper is devoted to an exact derivation of (\ref{A2 6}) in the atomic limit.
The wavenumber-dependent response function $\chiTk$ does not appear to be easily analyzed with the rigorous methods used in \cite{F1,CLY}. Here, we use the technique of quantum Mayer graphs that gives a straightforward expansion for two particle correlation functions needed to calculate $\chiTk$. It allows a derivation of \eqref{A2 6} that is exact in the sense that it does not involve any intermediate model or approximation. In particular the many-body problem is fully taken into account and the existence of preformed atoms is not assumed. However the derivation remains formal to the extent that results are established for each individual graph, without control of the convergence of the diagrammatic series. The present technique is also suited to explore the vicinity of the atomic limit and to systematically calculate corrections to the ideal gas behaviour. For instance, non-ideal contributions to the Saha equation of state are derived in \cite{8bis}.

In section~4 we briefly recall the loop representation of the Coulomb gas. This formalism arises from the Feynman-Kac path integral representation of the quantum mechanical Gibbs factor. Charges of the same type and same statistics are collected according to a permutation cycle of length $q$ $(q=1,2,\ldots$) into a random path (a Brownian path) called a $q$-loop (for a review and references, see~\cite{BM}). In terms of loops, the statistical averages are performed according to the same rules as in classical statistical mechanics. 
As a consequence the powerful method of Mayer graphs is available in the space of loops. Thus an ensemble of point quantum mechanical charges becomes isomorphic to a classical-like system of fluctuating multipoles.
Two points can then conveniently be made at this stage. First, the divergences due to the long range of the Coulomb potential can be cured by the introduction, via partial resummations, of an effective screened potential (which is the quantum analogue of the usual Debye potential). The properties of this potential are studied in details in~\cite{BMA}. Morover the formalism offers the possibility of an easy derivation of the response function since the rules of classical linear response apply. In particular, in the langage of charged loops, the perfect screening relation (\ref{A2 -1}) takes the same form as  the  Stillinger-Lovett rule for a classical system of structured ions (see (\ref{A2 SL})). 

The loop expansion needs to be properly reorganized to perform estimations in the atomic limit. The reason is that an element of the space of loops consists into a number of charges of the same species, therefore not capable to bind. By a diagrammatical reorganization, the loop expansion can be converted into the \emph{screened cluster expansion} (section~5). The details of this reorganization are presented in~\cite{b13}. In the latter expansion, the basic elements are all possible clusters of positive and negative charges, thus candidates for atomic or molecular recombination. Let us just mention that for a system of particles interacting via a short-range potential, this expansion exactly coincides with the usual quantum mechanical virial expansion. In the Coulomb case, the expansion undergoes a number of modifications because of the necessity to deal with an effective screened potential. 

At this point, we have in hand the necessary tools to investigate the response function in the atomic limit. In section~6, we select and study the leading graphs yielding the dielectric response. There are essentially two aspects to be controlled. As the density goes to zero, the screened potential reduces to the (non integrable) bare Coulomb potential. One must make sure that this does not create any divergence in the atomic limit. Secondly, one must control that all effects due to excited states and ionized states of the hydrogen atom become negligible when the temperature vanishes. We find it worth to give complete proofs of these non trivial mathematical points (technical parts of the proofs are relegated in appendix~A).
We check then that all the other graphs involving clusters of more than two particles do not contribute in the limit. For this we rely on the general analysis presented in~\cite{b13}. 
Eventually, section~7 is devoted to a discussion of the response function in the wave number region that interpolates between perfect and dielectric screening. Our results are compared with the extended RPA dielectric function introduced in the framework of standard many-body perturbation theory.

\section{The atomic limit}	\label{A2 S Atomic limit}

The notion of atom or molecule in the many-body problem can only receive a precise
meaning in an asymptotic sense. The temperature must tend to zero to give a predominant 
weight to bound states over ionized states and the density $\rho$ should be small enough
($\rho^{-1/3}\gg \aB$, the Bohr radius), to have spatially non overlapping atoms.
To formulate the atomic limit, we introduce first the grand canonical densities of three independent species, the electrons (e), the protons (p) and the hydrogen atoms (a) in their ground state,
\eq{ 
\rho_{j}^{\rm id}= d_j \parenth{\frac{m_j}{2\pi \beta
\hbar^2}}^{3/2}\exp(-\beta(E_j-\mu_j)),\;\;\,\;\;\;j=\text{e},\:\text{p},\:\text{a} \label{A2 2.1}
}
where $\mu_j$ are the respective chemical potentials, $m\atH=m\el+m\pr,\;\; E\el=E\pr=0$, $E\atH<0$ is the ground state energy of the hydrogen atom, and $d_j$ is the spin degeneracy ($d_{{\rm a}}=4$, $d\el = d\pr = 2$). Here all the effects of the Coulomb interaction are disregarded except for binding energy $|E\atH|$ of the atom.

The law of chemical equilibrium for the dissociation reaction $\text{e}+\text{p}\leftrightarrow \text{a}$ requires
\eq{
\mu\atH=\mu\el+\mu\pr
\label{A2 2.2}
}
and one also must have charge neutrality $\rho\el^{\rm id}=\rho\pr^{\rm id}$. Introducing the combinations 
\eq{
\mu=\frac{\mu\el+\mu\pr}{2},\;\;\,\nu=\frac{\mu\el-\mu\pr}{2}
\label{A2 2.4}
}
it is easily seen that the neutrality condition imposes the choice
\eq{
\nu=\nu(\beta)=\frac{3}{4\beta}\ln\frac{m\pr}{m\el} = \OO(\be^{-1}).
\label{A2 2.5}
}
Hence (\ref{A2 2.1}) becomes
\begin{align}	\label{density rho_el_id}
\rho\el^{\rm id} &= \rho\pr^{\rm id}=\frac{2}{(2\pi\lam\el\lam\pr)^{3/2}}\e{\beta\mu} \\
\rhoat &= \frac{4}{(2\pi \lambda\atH^{2})^{3/2}}\e{-\beta(E\atH-2\mu)}
\label{A2 def rho_at}
\end{align}
with
\eq{
\lam_\al=\hbar\sqrt{\frac{\beta}{m_\al}}
\label{A2 def lam_at}
}
the thermal de Broglie length of a particle of mass $m_\al$. The ideal density of an atom, molecule or ion with $N\pr$ protons and $N\el$ electrons and ground state energy $E_{N\pr N\el}$ is of the form
\eq{
\rho^{\rm id}_{N\pr N\el} = \frac{d_{N\pr N\el}}{(2\pi\lambda^{2}_{N\pr N\el})^{3/2}}
\exp[-\beta(E_{N\pr N\el}-\mu(N\el +N\pr)+ \OO(\beta^{-1})) ]
\label{A2 2.7}
}
where we used the fact that $\mu\el N\el + \mu\pr N\pr = \mu(N\el + N\pr) + \nu (N\el - N\pr)$ and (\ref{A2 2.5}) has been taken into account. $\lambda_{N\pr N\el}$ is the corresponding thermal length and $d_{N\pr N\el}$ the degeneracy factor.

The situation where the atomic density dominates all other ionic or molecular densities at low temperature is characterized by
\eq{
\rhoat\gg\rho^{\rm id}_{N\pr N\el},\;\;\; \beta\to \infty.
\label{A2 2.8}
}
More generally, we require that the probability of occurrence of hydrogen atoms dominates that of all possible other configurations of protons and electrons. This will happen if there exists a range~$I$ of chemical potential $\mu$
such that all the inequalities 
\eq{
0<E\atH-2\mu<E_{N\pr N\el}-\mu(N\el +N\pr),\;\;\;(N\el ,N\pr)\neq (0,0),\;(1,1)
\label{A2 2.9}
}
can be simultaneously satisfied, where $E_{N\pr N\el}$ is the infimum of the spectrum of $\opH_{N\pr N\el}$. The range~$I$ (if any) is located above $E\atH$ since setting $(N\el ,N\pr)=(1,0)$ (single electron) in (\ref{A2 2.9}) gives $\mu>E\atH$. Moreover setting $(N\el ,N\pr)=(2,2)$ (hydrogen molecule) implies
$\mu<\frac{1}{2}(E_{22}-E\atH)$. Note that $E\atH<\frac{1}{2}(E_{22}-E\atH)$ since the binding energy $|E_{22}-2 E\atH|$ gained by the formation of a hydrogen molecule is less than the binding energy of the hydrogen atom himself, a well known fact. These two cases do not exhaust all the constraints imposed by the inequalities (\ref{A2 2.9}). From now on we make the plausible, but yet unproven, hypothesis that there exists an interval $I_\Delta=]E\atH,E\atH+\Delta], \Delta>0$, such that (\ref{A2 2.9}) holds when $\mu\in I_\Delta$.

It is worthwhile to note that the validity of this hypothesis is equivalent to the possibility to find an optimal constant for the stability of matter estimate that we write here in the form
\eq{
H_{N\pr N\el}\geq -B(N\el +N\pr-1),\;\; (N\el ,N\pr)\neq (0,0),(1,1)
\label{A2 2.16}
}
for some positive constant $B$. It is conjectured (but not proven) that there exists a stability constant $B$ strictly less than $|E\atH|$ for all cases except of course for the hydrogen
atom itself. It is then easily checked that the hypothesis $B<|E\atH|, \;(N\el ,N\pr)\neq (0,0),\:(1,1)$ in (\ref{A2 2.16}) is equivalent to the possibility of finding the non void interval $I_\Delta$ such that all the inequalities (\ref{A2 2.9}) are satisfied, $\mu\in I_\Delta$, hence assuring the existence of the atomic phase.

Under this condition, it has been proven that~\cite{F1,F2,CLY}
\eq{
p(\beta,\mu)=\rhoat(1+\OO(e^{-c\beta})),\;\;\,\mu\in I_\Delta, \;\,\;\beta\to \infty 
\label{A2 2.10}
}
namely that the grand canonical pressure $p(\beta,\mu)$ verifies the equation of state of a perfect
gas of hydrogen atoms of density $\rhoat$ up to an exponentially small correction in $\beta$. 
This is the situation we will consider in this paper and is referred to as the atomic limit
$\mu\in I_\Delta,\;\;\beta\to \infty$. 

In general, collective screening effects are provided not only by free electrons and protons, but also by all types of more complex ions that have appreciable densities $\rho_{N\pr N\el}\ide$ for given values of the temperature and the chemical potential. In this paper, we study the interplay between dielectric and ionic screening in the situation when the latter is predominantly due to unbound electrons and protons. This will certainly be the case if $\mu$ is chosen sufficiently close to~$E\atH$, in a possibly smaller interval $I_\delta = ]E\atH,E\atH+
\delta] \subset I_\Delta$ so that
\eq{
   0 < E\atH - 2\mu < -\mu < E_{N\pr N\el}-\mu (N\pr + N\el)
   \qquad \quad \mu\in I_\delta
}
holds for all $(N\pr,N\el)\neq(0,0),(1,0),(0,1),(1,1)$. We consider hence an atomic phase under the conditions that the most probable entities in the system are, after hydrogen atoms, ionized electrons and protons, namely
\eq{	\label{Strong inequ}
   \rho\atH\ide \gg \rho\el\ide = \rho\pr\ide \gg \rho_{N\pr N\el}\ide, 
   \qquad (N\pr,N\el)\neq(0,0),(1,0),(0,1),(1,1).
}
In the quantum Mayer graphs expansion, the collective screening effects are embodied into the effective screened potential obtained by chain resummations (see section~5). Under the conditions just described, we take as intermediate chain points those corresponding to individual electrons and protons. Then the resulting screening length $\kappa^{-1}$ in~\eqref{A2 D-H} will only involve the densities of free charges~\eqref{density rho_el_id}. If it turns out, in an other regime, that collective screening effects mainly arise from other types of ions, another definition of the effective potential must be adopted in consequence.

\section{Elementary description of screening}	\label{S elementary}

In this section, we describe, in an elementary (and non rigorous) manner, the screening of a classical external charge by an e-p plasma in the atomic limit. This heuristic description is intended to serve as a gentle introduction to the calculation of the next sections, in which the response function is studied in full generality.

In the atomic limit, the e-p plasma approaches an ideal gas of hydrogen atoms of density~$\rhoat$. Under the influence of an external charge, the hydrogen atoms get polarized and screen partially the external charge. At low temperatures and low densities, one expects this effect to be described, at leading order, by the dielectric constant~\eqref{A2 6}. According to~\eqref{A2 2} and at linear order in~$\rhoat$, the system's response function should therefore approach
\eq{	\label{A2 chi = rhoat alpha}
    \chiTk \simeq -4\pi\rhoat\alphaH
}
in an appropriate range of wave numbers.

In view of~\eqref{A2 4}, let us make a straightforward low activity expansion of the imaginary time displaced charge charge correlation function~$S(\vk,s)$. This correlation is defined by
\eq{	\label{A2 def S(k,s)}
    S(\vk,s) = \bracket{\tilde{\op{c}}(\vk,s)\op{c}(\vz)}
}
where $\bracket{\ldots}$ is the grand canonical average. For a $N$~particle system evolving with the Hamiltonian~$\opH_N$
\eq{	\label{A2 10}
    \op{c}_N(\ver) = \sum_{i=1}^N e_{\al_i} \delta(\ver-\opr_i)
}
is the microscopic charge density operator, and
\eq{	\label{A2 11}
    \tilde{\op{c}}_N(\vk,s) = \e{s\opH_N} \Big( \sum_{i=1}^N e_{\al_i} \e{-i\vk\cdot\opr_i} \Big)
    \e{-s\opH_N}
}
is its Fourier transform at ``imaginary time''~$s$. In general, if $\opA_N$ are $N$-particle observables, the first terms of the low fugacity expansion of the grand-canonical average of~$\opA$, with
\begin{equation}
    \bracket{\opA} = \frac{1}{\Xi}\sum_{N=1}^\infty \frac{z^N}{N!} \Tr\acc{\opA_N \e{-\be\opH_N}}, \qquad
    \Xi = \sum_{N=0}^\infty \frac{z^N}{N!} \Tr\, \e{-\be\opH_N}
\end{equation}
are
\begin{equation}	\label{A2 viriel}
    \bracket{\opA} = z \Tr \opA_1 \e{-\be\opH_1}
    	+ z^2 \Big[
    	\frac{1}{2} \Tr\, A_2 \e{-\be \opH_2} - (\Tr A_1\e{-\be\opH_1}) (\Tr\, \e{-\be\opH_1})
    	 \Big]	+ \OO(z^3)
\end{equation}
In the e-p system, the one-particle and two-particles Hamiltonians are
\begin{gather}
    \opH_\al = \frac{\opp^2}{2 m_\al}, \qquad \al=\text{e,p}	\\
     \opH_{\al_1 \al_2} = \frac{\opp_1^2}{2 m_{\al_1}} + \frac{\opp_2^2}{2 m_{\al_2}}
    	+ e_{\al_1} e_{\al_2} V(\opr_1-\opr_2)
\end{gather}
It is understood in~\eqref{A2 viriel} that the traces include a summation on the particle species and are carried out on properly antisymmetrized electronic and protonic quantum states.

We apply~\eqref{A2 viriel} to~\eqref{A2 def S(k,s)} and are interested in extracting from it the contribution proportional to~$\rhoat$ (see \eqref{A2 def rho_at}), which comprises the factor $\exp[-\be \Ea]$ where $\Ea$ is the ground state energy of the hydrogen atom. Such a contribution must come from the second term of~\eqref{A2 viriel}, the only one involving the two particle Hamiltonian. This contribution is then
\eq{	\label{A2 12}
    \frac{1}{2} \sum_{\al_1,\al_2} z_{\al_1} z_{\al_2} \sum_{\sig^z_1,\sig^z_2} \Tr\, \e{-\be\opH_{\al_1\al_2}}
    	\tilde{\op{c}}_2(\vk,s) \op{c}_2(\vz)
}
where we have made the sum over species and spin explicit and the trace is now purely configurational ($z_\al = \exp[\be\mu_\al]$ is the fugacity of the particles of species~$\al$). Moreover, only the part of~\eqref{A2 12} pertaining to an electron-proton pair with Hamiltonian~$\opH_{\text{ep}}$ will involve the factor $\exp[-\be \Ea]$. So keeping only the terms $\al_1\neq\al_2$ in~\eqref{A2 12} gives with~\eqref{A2 10}, \eqref{A2 11} and $\mu\el + \mu\pr = 2\mu$
\eq{
    4 \e{2\be\mu} e^2 \Tr\Big\{
    \e{-(\be-s)\opH_{\text{ep}}} (\e{-i\vk\cdot\opr\pr} - \e{-i\vk\cdot\opr\el}) \e{-s \opH_{\text{ep}}} (\delta(\opr\pr) - \delta(\opr\el))
    \Big\}.
}
We now proceed to the evaluation of this trace, going to the center of mass variables. Since
\eq{	\label{A2 def Hep}
    \opH_{\text{ep}} = \frac{\opp\el^2}{2 m\el} + \frac{\opp\pr^2}{2 m\pr} -e^2 V(\opr\el-\opr\pr)
}
separates into a center of mass Hamiltonian $\opH_{\text{CM}} = \vect{\op{P}}^2/2M$ and a relative Hamiltonian $\opH=\opp^2/2m - e^2/|\opr|$, its eigenstates are of the form $\ket{\vp,\n}=\ket{\vp}\otimes\ket{\n}$, with eigenvalue $E_{\vp} + E_{\n}$, where 
\eq{
    \opH_{\text{CM}} \ket{\vp} = E_{\vp} \ket{\vp}
    \qquad \text{and} \qquad
    \opH\ket{\n} = E_{\n} \ket{\n}
}
($\n$ is a generic index for the eigenstates (including the ionized states), $\n=0$ being the ground state). This allows one to write the e-p contribution to $\TF{S}(\vk,s)$ as
\begin{multline}	\label{A2 anon1}
    4\e{2\be\mu} e^2 \int\romd\vp\int\romd\vp' \sum_{\n,\n'\geq 0}
    	\e{-(\be-s)[E_{\vp} + E_{\n}]} \e{-s[E_{\vp'} + E_{\n'}]}
    	\bra{\vp,\n} \e{-i\vk\cdot\opr\pr} - \e{-i\vk\cdot\opr\el} \ket{\vp',\n'} \\ \times
    	\bra{\vp',\n'} \delta(\opr\pr) - \delta(\opr\el) \ket{\vp,\n}
\end{multline}
where $\opr\pr = \opR - \frac{m\el}{M}\opr$ and $\opr\el = \opR + \frac{m\pr}{M}\opr$ are the quantum operators associated to the position of the proton and of the electron. The eigenstates of $\opH_{\text{CM}}$ are of course plane waves $\bracket{\R|\vp}=(2\pi)^{-3/2}\exp[i\vp\cdot\R]$ with energy $E_{\vp} = \hbar^2\vp^2/2M$. The matrix elements in~\eqref{A2 anon1} are therefore equal to
\begin{gather}
    \bra{\vp,\n} \e{-i\vk\cdot\opr\pr} - \e{-i\vk\cdot\opr\el}\ket{\vp',\n'}
    	= \delta(\vp'-\vp-\vk) A_{\n\n'}(\vk)	\\
    \bra{\vp'=\vp+\vk,\n'} \delta(\opr\pr) - \delta(\opr\el) \ket{\vp,\n}
    	= \frac{1}{(2\pi)^3} A_{\n'\n}(-\vk)
\end{gather}
where we have defined
\eq{	\label{A2 def A_nn'}
    A_{\n\n'}(\vk) = \bra{\n} \e{i\vk\cdot\frac{m\el}{M}\opr} - \e{-i\vk\cdot\frac{m\pr}{M}\opr} \ket{\n'}.
}
From the above matrix elements, the e-p contribution to $\chiTk$ becomes
\eq{	\label{A2 anon2}
    \TF{\chi}_{\text{ep}}(\vk) \equiv -\frac{4\pi e^2}{k^2} 4\e{2\be\mu} \int_0^\be\romd \tau\,  f_{\tau/\be}(\vk) \sum_{\n,\n'\geq0}
    	\e{-(\be-\tau) E_{\n}} \e{-\tau E_{\n'}} |A_{\n\n'}(\vk)|^2.
}
where
\eq{	\label{A2 int dp}
    f_s(\vk) \equiv \frac{1}{(2\pi)^3} \int\romd\vp\, \e{-\be(1-s) E_{\vp}} \e{-\be s E_{\vp+\vk}}.
}
The Gaussian integral in $f_s(\vk)$ can be calculated with the result
\eq{	\label{A2 f_tau(k)}
    f_s(\vk) = \frac{1}{(2\pi\lamat^2)^{3/2}} \e{-\be E_{\vk} s(1-s)}
    = \frac{1}{(2\pi\lamat^2)^{3/2}} \e{-\frac{1}{2}k^2 \lamat^2 s(1-s)}
}
where $\lamat$ is the thermal wave length of the atom.
In the atomic limit, $\be$ is very large and the dominant terms in~\eqref{A2 anon2} are those with $\n=0$ or $\n'=0$, because they contain the exponentially growing factor $\exp[-\be E_0]$, which is much greater than $\exp[-\be E_{\n}]$ if $\n>0$ ($E_0=\Ea$). Keeping only these ground state terms and factoring out the atomic density~\eqref{A2 def rho_at}, we find that the hydrogen atoms give the dominant contribution
\eq{	\label{A2 chi_ep}
    \TF{\chi}_{\text{ep}}(\vk) \simeq -4\pi\rhoat \alphaH(\vk,\beta), \qquad \be\tend\infty,
}
to the response function, where
\eq{	\label{A2 al_be(k)}
    \alphaH(\vk,\beta) = \frac{e^2}{k^2} \int_0^\be\romd \tau\, \e{-\frac{1}{2}k^2\lamat^2 \frac{\tau}{\be}(1-\frac{\tau}{\be})}
    	\sum_{\substack{\n,\n'\geq0 \\ \n\cdot\n'=0}} \e{-\be(E_{\n}-E_0)}
    		\e{-\tau (E_{\n'}-E_{\n})} |A_{\n\n'}(\vk)|^2.
}
The quantity $\alphaH(\vk,\beta)$ can be interpreted as the polarizability of a hydrogen atom at inverse temperature $\be \gg (E_1-E_0)^{-1}$ in an external electric field varying on the scale~$k^{-1}$. Let us discuss briefly the result~\eqref{A2 al_be(k)} for $\alphaH(\vk,\beta)$, before commenting on the method used to derive it.

When $k \tend 0$, the external perturbation becomes uniform on larger and larger distances, so that one expects to recover the polarizability~$\alphaH$ of a hydrogen atom in an uniform electric field. For this, it is necessary to have the perturbation uniform on the scale of the dispersion of the center of mass distribution, namely $k\ll\lamat^{-1}$. If it is the case, the factor $\exp[-\frac{1}{2}k^2\lamat^2 \frac{\tau}{\be}(1-\frac{\tau}{\be})]$ in~\eqref{A2 al_be(k)} can be approximated by one and the $\tau$-integral in~\eqref{A2 al_be(k)} is easily evaluated:
\eq{	\label{A2 47}
    \alphaH(\vk,\beta) \simeq \frac{2 e^2}{k^2}\sum_{\n=0}^\infty \frac{1-\e{-\be(E_{\n}-E_0)}}{E_{\n}-E_0} |A_{0\n}(\vk)|^2, \qquad k\lamat \ll 1.
}
In obtaining~\eqref{A2 47}, we used the symmetry properties of the matrix element: $A_{\n\n'}(\vk) = A_{\n' \n}(-\vk)$ and $|A_{\n\n'}(\vk)|^2=|A_{\n\n'}(-\vk)|^2$ (this follows from the fact that $\opH$ commutes with the parity operation).
Moreover, for $\be$~large, $\lamat$ is much greater than the Bohr radius, and hence $k\ll \lamat^{-1}\ll\ab^{-1}$, so that one can make the dipole approximation
\eq{	\label{A2 dipole approx}
    |A_{0\n}(\vk)|^2 \simeq k^2 |\bra{0} \hat{\vk}\cdot\opr\ket{\n}|^2, \qquad k\ll \ab^{-1}.
}
Notice that $\bra{0}\opr\ket{0}=0$ by parity. For $k\ll\lam\atH^{-1}$ and $\be \gg (E_1-E_0)^{-1}$, $\alphaH(\vk,\beta)$ reduces therefore indeed to the ground state polarizability of a hydrogen atom in an uniform electric field~\cite{Landau}:
\eq{    \label{A2 formule alpha}
    \alphaH(\vk,\beta) \simeq \alphaH =
     2 e^2 \sum_{\n \geq 1} \frac{\absol{\bra{0} \hat{\vk}\cdot\opr \ket{\n}}^2}{E_{\n} - E_0} = \frac{9}{2} \ab^3,\qquad k\ll\lamat^{-1}\ll\ab^{-1}.
}
(The term $n=0$ in~\eqref{A2 47} is of order $\be e^2 (k \aB)^4/k^2=\aB^3(k \be e^2)(k \aB)$ and can be neglected when $k \ll \lam\atH^{-1}$.) This will give the anticipated form~\eqref{A2 chi = rhoat alpha} of the response function provided that $k$ is not in a range where perfect screening~\eqref{A2 -1} is prevailing. In view of~\eqref{A2 D-H}, we must require
\eq{	\label{A2 kap}
    \frac{\kappa^2}{k^2+\kappa^2} \ll 4\pi\rhoat\alphaH,
}
or, equivalently,
\newcommand{\lami}{\lam_{\text{I}}}
\eq{	\label{A2 lami}
    k^2 \gg \frac{\be e^2}{\alphaH} \frac{\rho\el^{\rm id}}{\rhoat} \equiv \lami^{-2}.
}
In~\eqref{A2 kap},
\eq{	\label{A2 def kappa2}
    \kappa = \sqrt{8\pi\be e^2 \rho\el^{\rm id}}
}
is the usual inverse Debye length for an e-p plasma and Eq.~\eqref{A2 lami} defines a length~$\lami$ which is the border-line scale below which ionic screening starts taking place.

Summarizing the discussion, we expect that in the vicinity of the atomic limit, the response function of the e-p system should be essentially constant and equal to the value~\eqref{A2 chi = rhoat alpha} in the range
\eq{	\label{A2 plateau}
    \lami^{-1} \ll k \ll \lamat^{-1}.
}
This region corresponds to the dielectric behaviour of the e-p system in its atomic phase. As $k$~is further decreased, ionic screening will predominate, and $\chiTk$ interpolates between the values $-4\pi\rhoat\alphaH$ and $-1$. Notice finally that $\chiTk \tend 0$ as $k\tend\infty$ since atoms cannot be polarized under the effect of an electric field with infinitely fast spatial oscillations.

The cross-over length $\lam_{\rm I}$ between dielectric and ionic screening was derived in an infinitely extended state of the e-p system. The same length can also be obtained by considering the screening of a point external charge $e_0$ immersed in a spherical sample of radius~$L$. We assume $L$ much smaller than the Debye screening length ($L\ll \kappa^{-1}$). The charge density of free charges induced around $e_0$ is given in the Debye-H\"{u}ckel approximation by $c_{\text{free}}(\ver) = - e_0 \kappa^2 \exp(-\kappa r)/(4\pi r)$. The screening provided by the free charges is hence
\eq{
    \int_{|\ver|<L} \romd \ver\, c_{\text{free}}(\ver) = - e_0 \int_0^{\kappa L} \romd u\,
      u \e{-u} \simeq - e_0 \frac{1}{2}(\kappa L)^2.
}
Comparing with the induced charge due to polarisation $- 4\pi\rho\atH\ide \alphaH e_0 $, the sample is indeed expected to display a dielectric behaviour if $L^2 \ll 8\pi\rho\atH\ide \alphaH / \kappa^2 = \lam_{\rm I}^2$, in agreement with the cross-over distance defined in~\eqref{A2 lami}.\\

The present elementary description of screening shows that the dielectric regime of the response function is formally characterized by
\eq{	\label{A2 THM}
    \lim_{\substack{\be\tend\infty\\k\tend 0}} \frac{\chiTk}{\rhoat} \tend -4\pi\alphaH,\qquad k\lamat\tend 0, \; k\lami\tend\infty, \; \mu \in I.
}
This should be contrasted with a strict zero temperature limit at fixed $\vk$. If one takes the limit $\be\tend\infty$ of $\alphaH(\vk,\beta)$ at fixed $\vk$ (let $s=\tau/\be$ in~\eqref{A2 al_be(k)} and note that $\frac{1}{2}k^2 \lamat^2 = \be E_{\vk}$), a Dirac delta function appears:
\begin{multline}
    \lim_{\be \tend \infty} \be \e{-\be[E_{\n} - E_0 + s(E_{\n'} - E_{\n} -E_{\vk}) - E_{\vk}s^2]}
    = 2 \delta(E_{\n} - E_0 + s(E_{\n'} - E_{\n} -E_{\vk}) - E_{\vk}s^2) \\
    = \frac{2 \delta_{\n,0}}{E_{\vk} + E_{\n'} - E_0} \delta(s)
    	+\frac{2 \delta_{\n',0}}{E_{\vk} + E_{\n} - E_0} \delta(s-1), \qquad \n\cdot \n'=0.
\end{multline}
Using $\int_0^1\romd s\, \delta(s) = 1/2$ and the symmetry properties of the matrix element $|A_{0\n}(\vk)|$, we find
\eq{	\label{A2 anon3}
   \lim_{\be\tend\infty}\alphaH(\vk,\beta) = \frac{2 e^2}{k^2} \sum_{\n\geq 0}
    	\frac{\absol{A_{0\n}(\vk)}^2}{E_{\vk} + E_{\n} - E_0}
    	=  \alphaH(\vk,\infty).
}
If we now let $k \tend 0$ in~\eqref{A2 anon3}, we obtain
\eq{	\label{A2 anon4}
    \lim_{k \tend 0}  \alphaH(\vk,\infty) = 
    \alphaH + \frac{e^2 (m\pr^2 - m\el^2)^2}{9 \hbar^2 M^3} |\bra{0}\opr^2\ket{0}|^2
}
which is obviously different from~\eqref{A2 formule alpha}. The result~\eqref{A2 anon4} can also be obtained from a direct calculation in the ground state of~$\opH_{\text{ep}}$ \eqref{A2 def Hep}. The additional contribution in~\eqref{A2 anon4} has its origin in the coupling of the center of mass of the atom with the external electric field. This contribution does not vanish in the limit $k \tend 0$, because the external field does not appear to be uniform (even when $k$ is very small) to the atoms, the latter being entirely delocalized at zero temperature (here $k^{-1}\ll\lamat=\infty$).\\

We comment now on the above calculation of $\chiTk$: in fact all steps of this calculation are ill-defined, except for the final result~\eqref{A2 chi_ep} and~\eqref{A2 al_be(k)}. First of all, since $\chiTk$ represents the response to an external charge density localized in the bulk, the infinite volume limit should be taken first to disregard all boundary effects. But for a non-integrable potential, all virial coefficients of the series~\eqref{A2 viriel} (except the first one) diverge in the thermodynamic limit. The first term in~\eqref{A2 viriel} gives the contribution (the factor~2 accounts for the spin degeneracy)
\eq{	\label{A2 -kappa^2(k)/k^2}
    -\frac{4\pi\be}{k^2} 2 \sum_\al e_\al^2  z_\al \int_0^1\romd s\,
    	\bra{\vz}\e{s\opH_\al} \e{-i\vk\cdot\opr} \e{-s \opH_\al} \ket{\vz}
    = - \frac{\kappa^2(k)}{k^2}
}
to the response function, where we defined the function
\eq{	\label{A2 def kappa2(k)}
    \kappa^2(k) \equiv 4\pi\be \sum_{\al} e_{\al}^2 \frac{2 z_\al}{(2\pi\lam_\al^2)^{3/2}}
    	\int_0^1\romd s\, \e{-\frac{1}{2}k^2 \lam_\al^2 s(1-s)}.
}
Notice that~$\kappa^2(k)$ reduces at $k=0$ to $\kappa^2$ (see~\eqref{A2 def kappa2}), using the definitions~\eqref{A2 2.1} of the ideal densities and the neutrality condition $\rho\pr\ide=\rho\el\ide$. Hence, when $k\lam_\al\ll 1$, $\kappa^2(k)\simeq \kappa^2$, and \eqref{A2 -kappa^2(k)/k^2} is essentially identical to~\eqref{A2 D-H} as long as $k \gg \kappa$. However, since~\eqref{A2 -kappa^2(k)/k^2} diverges when $k \tend 0$, while $\chiTk$ goes to~$-1$ in this limit, there must obviously exist some other contributions, coming from $n$-body states in the series~\eqref{A2 viriel}, which cannot be neglected even at low density. Physically, these many-body contributions are associated to a collective effect in the system: the screening of the Coulomb interaction by ``screening clouds'' of particles appearing around the charges. This effect must be taken into account systematically in order to cure the problem of the Coulomb divergences in~\eqref{A2 viriel}.

Furthermore, besides the collective screening effects, we have not dealt with the more complex entities (e.g. the hydrogen molecule) that can be formed by chemical binding. In view of \eqref{A2 2.8} and~\eqref{A2 2.9}, such entities should not contribute to the dielectric response in the regime defined by the condition~\eqref{A2 plateau} in the atomic limit. Finally, one should check that in this limit, excited and ionized states of the hydrogen atom do not contribute either. The main goal of the following sections is to provide a calculation algorithm which is free from these difficulties.

\section{The loop representation of~$\chiTk$}	\label{S loop representation}

We recall in this section the loop representation of the response function~$\chiTk$. That representation is obtained when the grand partition function~$\Xi_\Lambda$ of the quantum plasma is written in a classical form, by using the Feynman-Kac path integral formula~\cite{SimSchRoe} and collecting permutations with the same cycle structures~\cite{Ginibre,Cornu96}:
\eq{	\label{A2 magic}
    \Xi_\Lambda = \sum_{N=0}^\infty \frac{1}{N!} \int \prod_{i=1}^{N} \romd\LL_i\,
    	z(\LL_i) \e{-\be U(\LL_1,\ldots,\LL_N)}.
 }
This so-called magic formula relies on the following definitions.
The element of phase space~$\LL$, called a loop, is a collection of $q$ particles of the same species exchanged in a cycle. A loop
\eq{
    \LL = (\R,\al,q,\X(s)), \qquad 0 \leq s \leq q
}
is specified by its position $\R$ in space, a particle species~$\al$, a number of particles~$q$, and a shape $\X(s)$ with $\X(0) = \X(q) = \vz$. The closed path
\eq{
   \R(s) = \R + \lam_\al \X(s)
}
describes the trajectories of the $q$ particles, which are located at $\R(k-1), k=1,...,q$ as they move from their position to the position of the next particle in the cycle in unit ``time''. The path $\X(s)$ is distributed according to a normalized Gaussian measure $\romD(\X)$, with covariance
\eq{	\label{A2 cov}
    \int \romD(\X) \X_\mu(s)\X_\nu(t) = \delta_{\mu,\nu} q \Big[\min\Big(\frac{s}{q},\frac{t}{q}\Big)-\frac{st}{q^2}\Big].
}
Integration over phase space means integration over space and summation over all internal degrees of freedom of the loop (which we denote collectively by $\chi=(\al,q,\X)$:
\eq{
    \int\romd\LL \cdots= \int\romd\R\int\romd\chi \cdots= \int\rom\romd\R \sum_{\al=1}^{\cal S} \sum_{q=1}^\infty \int\romD(\X) \cdots
}
The interaction energy of $N$ loops is the sum of two-body interaction potentials $U(\LL_1,...,\LL_N) = \sum_{1=i<j}^N V(\LL_i,\LL_j)$
with the interaction between two different loops
\eq{	\label{A2 def V(LL_i,LL_j)}
    V(\LL_i,\LL_j) = e_{\al_i}e_{\al_j}\int_0^{q_i}\romd s \int_0^{q_j}\romd t \, \TF{\delta}({s}-{t})\,
    	V(\R_i(s) - \R_j(t)).
}
In \eqref{A2 def V(LL_i,LL_j)}, $\TF{\delta}(s)=\sum_{n=-\infty}^\infty \delta(s-n)$ is the Dirac comb of period one. $V(\LL_i,\LL_j)$ is hence the sum of the Coulomb interactions between the particles in the loop $\LL_i$ and the particles in the loop $\LL_j$ as they move along their trajectory. The loop potential is clearly a function of the relative distance $\R_i - \R_j$ and of the internal constitution of the loops:
\eq{
    V(\LL_i,\LL_j) = V(\R_i - \R_j,\chi_i,\chi_j).
}
Eventually, the activity of a loop reads
\eq{	\label{A2 def z(LL)}
    z(\LL) = (-1)^{q-1} \frac{2}{q} {\frac{z^q_\al}{(2\pi q \lam_\al^2)^{3/2}}}
    	\e{-\be U(\LL)}, \qquad z_\al = \e{\be\mu_\al}
}
(the factor 2 takes the spin degeneracy into account) where
\eq{
    U(\LL) = \frac{e_\al^2}{2} \int_0^q\romd s_1 \int_0^q \romd s_2\, (1-\delta_{[s_1],[s_2]})
    	\TF{\delta}({s}_1-{s}_2) V(\R(s_1)-\R(s_2))
}
is the sum of the mutual interactions of the particles within a loop (the factor $(1-\delta_{[s_1],[s_2]})$ excludes the self-energies of the $q$~particles). The above rules define the statistical mechanics of the system of charged loops, which we call the loop representation of the quantum plasma.
Note that the interaction potential~\eqref{A2 def V(LL_i,LL_j)} inherited from the Feynman-Kac formula is not equal to the electrostatic interaction between two classical charged wires, which would be
\eq{	\label{A2 def V_elec(LL_i,LL_j)}
    V_{\text{elec}}(\LL_i,\LL_j) = e_{\al_i}e_{\al_j}\int_0^{q_i}\romd s \int_0^{q_j}\romd t \,V(\R_i(s) - \R_j(t))
}
Although the formalism of loops has a classical structure, the difference between $V(\LL_i,\LL_j)$ \eqref{A2 def V(LL_i,LL_j)} and $V_{\text{elec}}(\LL_i,\LL_j)$ is responsible for the absence of exponential screening in the quantum plasma~\cite{AM}. This difference is the occurrence of the equal time condition $\TF{\delta}({s}-{t})$ which characterizes the quantum mechanical aspect of the interaction~\eqref{A2 def V(LL_i,LL_j)}.

In the loop representation, one can define the loop correlation functions according to the usual definitions. Introducing the loop density $\hat{\rho}(\LL)=\sum_i\delta(\LL,\LL_i)$, the average loop density and the two-loop distribution function are
\eq{
    \rho(\LL)=\bracket{\hat{\rho}(\LL)}, \qquad \rho_{\TT}(\LL_1,\LL_2) = \bracket{\hat{\rho}(\LL_1)\hat{\rho}(\LL_2)} - 
    \bracket{\hat{\rho}(\LL_1)} \bracket{\hat{\rho}(\LL_2)},
}
where the average is taken with respect to the statistical ensemble of loops defined in~\eqref{A2 magic}, and coincident points are included in~$\rho_{\TT}(\LL_a,\LL_b)$. It is appropriate to recall that the charge sum rule holds in the system of loops~\cite{BMA}
\eq{	\label{A2 CSR loop}
    \int \romd \ver\int\romd\chi_a \, q_a e_{\al_a} \,\rho_{\TT}(\ver,\chia,\chib) = 0.
}
Any fixed loop of charge~$q_b e_{\al_b}$ is surrounded by a cloud of loops of opposite total average charge.

When a localized external charge~$c_{\text{ext}}(\ver)$ is immersed in the plasma, the corresponding induced charge density~$c_{\text{ind}}(\ver)$ can be calculated, within the formalism of loops, according to the rules of classical linear response theory. The external potential
\eq{	\label{A2 74bis}
    V_{\text{ext}}(\ver)=\int\romd \ver' \frac{c_{\text{ext}}(\ver')}{|\ver-\ver'|}
}
due to $c_{\text{ext}}(\ver)$ is represented in the system of loops by
\eq{	\label{A2 pot ext.}
    V_{\text{ext}}(\LL) = \int_0^q \! \romd s \, V_{\text{ext}}(\R + \lam_\al \X(s)).
}
It gives rise to an interaction energy
\eq{	\label{A2 U_ext loop}
    U_{\text{ext}} = \int\romd \LL \, e_{\al} \hat{\rho}(\LL) V_{\text{ext}}(\LL)
}
which has to be added to the pair interaction of loops~$U$ in~\eqref{A2 magic}. Then the response of the loop density~$\rho_{\text{ind}}(\LL)$ in presence of the external potential~\eqref{A2 pot ext.} is given at linear order by the standard formula
\eq{
    \rho_{\text{ind}}(\LL) = -\be \int\romd \LL' \, e_{\al'} V_{\text{ext}}(\LL') \rho_{\TT}(\LL,\LL')
}
To form the induced charge density at~$\ver$, one has to integrate~$\rho_{\text{ind}}(\LL)$ on the shape of the loop and take into account that a loop carries total charge~$e_\al q$:
\eq{	\label{A2 c_ind(r) loop}
    c_{\text{ind}}(\ver) = -\be \int\romd \R'\int\romd\chi\int\romd\chi'\,
    	e_{\al} q e_{\al'} \int_0^{q'}\! \romd s\, V_{\text{ext}}(\R' + \lam_{\al'}\X'(s))
    		\rho_{\TT}(\ver,\chi,\R',\chi').
}
The final expression for the dimensionless response function~\eqref{A2 def chi} follows after Fourier transformation of~\eqref{A2 c_ind(r) loop}, taking~\eqref{A2 74bis} into account,
\begin{equation}	\label{A2 chiTk loop}
    \chiTk = -\frac{4\pi\be}{k^2} \int\romd\chi_a\int\romd\chi_b\, e_{\al_a}q_a e_{\al_b} \int_0^{q_b} \romd \tau_b  \, \e{i\vk\cdot\lam_b\X_b(\tau_b)}
    	\TF{\rho}_{\TT}(\vk,\chi_a,\chi_b).
\end{equation}
Here $\TF{\rho}_{\TT}(\vk,\chi_a,\chi_b)$ is the Fourier transform of the translation invariant truncated loop-loop density fluctuation~$\rho_{\TT}(\ver,\chi_a,\chi_b)$. The formula~\eqref{A2  chiTk loop} has first been derived by Cornu~\cite{Cornu96}.\\
%She proved in particular in the framework of the loop formalism the perfect screening %relation~\eqref{A2 -1}.\\

There is an interesting interpretation of the perfect screening relation~\eqref{A2 -1} in terms of the statistical mechanics of random charged loops. We split~$\chiTk$ into two parts
\eq{
    \chiTk = -\frac{4\pi\be}{k^2} (\TF{S}^{\text{loop}}(\vk) + \TF{M}^{\text{loop}}(\vk))
}
where
\eq{
    \TF{S}^{\text{loop}}(\vk) = \int\romd\chi_a\int\romd\chi_b\, e_{\al_a} q_a e_{\al_b} q_b
    \,	\TF{\rho}_{\TT}(\vk,\chi_a,\chi_b)
}
is the Fourier transform of the charge-charge correlation of loops and
\eq{
    \TF{M}^{\text{loop}}(\vk) = \int\romd\chi_a\int\romd \chi_b\, e_{\al_a} q_a e_{\al_b} \int_0^{q_b}\romd\tau (\e{i\vk\cdot\lam_{\al_b}\X_b(\tau)}-1) \, \TF{\rho}_{\TT}(\vk,\chi_a,\chi_b)
}
comprises the multipolar contributions of the loops to~$\chiTk$. Because of the charge sum rule~\eqref{A2 CSR loop} together with rotational invariance, both~$\TF{S}^{\text{loop}}(\vk)$ and $\TF{M}^{\text{loop}}(\vk)$ have to be~$\OO(|\vk|^2)$ as $k \tend 0$. The rotational symmetry forces these~$|\vk|^2$ terms to take the form
\eq{
    \TF{S}^{\text{loop}}(\vk) \sim |\vk|^2 \frac{1}{6} \int\romd\ver\, |\ver|^2 S^{\text{loop}}(\ver),
    	\qquad k \tend 0
}
and
\begin{multline}	\label{A2 prev}
    \TF{M}^{\text{loop}}(\vk) \sim i \int\romd\chi_a\romd\chi_b \, e_{\al_a} q_a	(\vk\cdot\vect{d}(\chi_b))(\vk\cdot\grad_{\vk}\,\TF{\rho}_{\TT}(\vk,\chi_a,\chi_b)\big|_{k=0})
    	\\
    = \frac{|\vk|^2}{3} \int\romd\ver\, \ver\cdot\vP^{\text{loop}}(\ver)
    = - \frac{|\vk|^2}{6} \int\romd\ver\,|\ver|^2 \grad\cdot\vP^{\text{loop}}(\ver),
    	\qquad k \tend 0.
\end{multline}
In~\eqref{A2 prev}, we have defined the dipole of a loop by
$
    \vect{d}(\chi) = e_\al q \lam_\al \int_0^q\romd\tau\, \X(\tau)
$
and introduced the polarization vector
\eq{
    \vP^{\text{loop}}(\ver) = \int\romd\chi_a\int\romd\chi_b\, e_{\al_a} q_a \vect{d}(\chi_b)
    	\rho_{\TT}(\ver,\chi_a,\chi_b).
}
as equal to the charge-dipole correlation of loops. With these definitions, the perfect screening relation written in terms of charge-charge and charge-dipole correlation of loops takes the classical form of the second moment Stillinger-Lovett condition, namely
\eq{	\label{A2 SL}
    \frac{4\pi \be}{6} \int\romd\ver\, |\ver|^2 (S^{\text{loop}}(\ver) - \grad\cdot\vP^{\text{loop}}(\ver)) = -1.
}
The same relation holds in classical models of structured ions where both the charge density and the polarization charge~$-\grad\cdot\vP(\ver)$ participate to the constitution of the screening cloud~\cite{MartinGruber}.

\section{The screened cluster expansion}	\label{S SCE}

The loop formalism leads itself naturally to the introduction of Mayer graphs on the space of loops. A vertex receives the weight~$z(\LL)$~\eqref{A2 def z(LL)} and a bond the factor $\exp[-\be V(\LL_i,\LL_j)] - 1$. Since the loop pair potential~\eqref{A2 def V(LL_i,LL_j)} behaves as the Coulomb potential $q_i e_{\al_i} q_j e_{\al_j} / |\R_i-\R_j|$ itself, the bonds are not integrable at large distances, and partial resummations are needed. All divergencies can be removed by introducing the effective screened potential $\phi(\LL_a,\LL_b) = \phi(\R_a-\R_b,\chia,\chib)$ defined as the sum of chains
\eq{	\label{A2 def graph phi}
-\be\phi(\LL_i,\LL_j) \equiv \raisebox{1mm}{\Graph{Fig_Fc2.ps}}
}
where the bond is the linearized Mayer bond $-\be V(\LL_i,\LL_j)$. This potential is the quantum analogue of the classical Debye potential (see~\cite{BMA}). For the purpose of this paper, it is convenient to use a slightly different definition of $\phi$ by restricting the intermediate (black) points to one-particle loops. Its properties are the same as those of the effective potential studied in~\cite{BMA}. The potential $\phi$ describes the screening effects due to ionized protons and electrons with the inverse Debye screening length $\kappa$ \eqref{A2 kap}. At short distances $|\ver|\ll\kappa^{-1}$, $\phi$ reduces to the bare Coulomb potential $V(\LL_a,\LL_b)$ between loops. At distances $|\ver|\sim \kappa^{-1}$, $\phi \simeq q_a e_{\al_a} q_b e_{\al_b} \exp[-\kappa r]/r$ approaches the standard Debye potential that describes the classical collective screening effects. At large distances, $\phi$ has an $1/r^{3}$ tail corresponding to dipole-dipole interactions between the loops (this tail is responsible for the algebraic decays of the correlations in the quantum plasma).

Once all chain summations \eqref{A2 def graph phi} are performed in the Mayer graphs, the resulting prototype graphs still obey Mayer diagrammatic rules, with two kinds of bonds ($\FC=-\be\phi$ and $\FR=\exp[-\be\phi] - 1 + \be\phi$), two kinds of weights, and two additional rules that prevent double counting (see~\cite{BMA}). Like in a system of classical dipoles, the bond $\FC(\LL_a,\LL_b)$ is at the borderline of integrability. The prototype graphs are integrable at large distances provided that the integrations on the internal variables of the loops (the shape~$\X(s)$) are performed first.

Though the prototype graphs describe non perturbatively screened Coulomb interactions between quantum particles, they are not adapted to the evaluation of the response function $\chiTk$ in the atomic limit. Indeed, the small parameters in this limit are the ideal densities (section~\ref{A2 S Atomic limit})
\eq{
   \rho_{N\pr N\el}\ide \propto \e{-\be(E_{N\pr N\el} -\mu(N\pr + N\el))} \ll 1
}
which are not apparent (or even present) in individual prototype graphs. This problem occurs because the vertices in prototype graphs involve single loops, which are groups of particles of the same species, with only repulsive pair-wise interactions (included in the loop-activity~\eqref{A2 def z(LL)}). Attractive interactions appear in the graphs as bonds connecting electronic and protonic loops, but a given graph does not in general include the total set of pair-wise interactions that would be necessary to form bound states between electrons and protons (see~\cite{b13}). Individual loop-Mayer graphs are therefore not in direct correspondence with the ideal atomic or molecular densities occurring in the atomic limit. For this reason, it is convenient to collect prototype graphs to form new graphs involving clusters of protons and electrons, together with all their mutual interactions and proper statistics, in such a way that the new effective activities approach the ideal densities in the atomic limit. This reorganization, called the screened cluster expansion, is worked out in details in~\cite{b13}. We recall here only the final diagrammatic rules, taking into account the minor modifications introduced by the choice of a slightly different effective potential~$\phi$. We stress that this expansion is nothing but the usual quantum cluster expansion, suitably generalized to take into account the screening due to the long range of the Coulomb potential.\\

The screened cluster expansion for equilibrium quantities of the quantum e-p plasma is expressed in terms of Mayer graphs $G$, with the following definitions of bonds and weights~\cite{b13}.

\paragraph{Vertices}
A vertex $C$ in a graph $G$ is a cluster of particles, denoted by $C(N\pr,N\el)$, containing $N\pr$ protons and $N\el$ electrons. The internal state of a cluster involves all possible partitions of the protons and electrons into sets of protonic and electronic loops. Let
\eq{
    Q_\al=[q_1,\ldots,q_{L_\al}], \qquad \sum_{i=1}^{L_\al} q_i = N_\al
}
be a partition of $N_\al$ into $L_\al$~subsets of $q_k$~particles, $k=1,\ldots,L_\al$, with $q_1\geq q_2 \geq \ldots \geq q_{L_\al}$. Here $L_\al$ runs from 1 to~$N_\al$. To a partition $(Q\pr,Q\el)$ of the $N\pr$ protons and $N\el$ electrons, we associate a cluster of loops
\eq{
    \CC(Q\pr,Q\el) = \{ \LL_1^{(\rmp)},...,\LL_{L\pr}^{(\rmp)},\LL_1^{(\rme)},...,\LL_{L\el}^{(\rme)} \}
}
where $\LL_k^{(\al)}$ carries $q_k^{(\al)}$ particles of species~$\al$ ($k=1,...,L_\al$).
The variables associated to a cluster $C(N\pr,N\el)$ of a graph $G$ are $Q\pr,Q\el,\CC(Q\pr,Q\el)$. The statistical weight of a cluster reads
\begin{equation} 	\label{A2 Z(C)}
Z_{\phi}^{\TT}(C) = {\prod_{k=1}^{L_p} z_{\phi}({\cal L}_k^{(\rmp)}) 
\prod_{k=1}^{L\el} z_{\phi}({\cal L}_k^{(\rme)}) \over 
\prod_{q=1}^{N\pr} n\pr(q)! \prod_{q=1}^{N\el} n\el(q)!} 
{\cal B}_{\phi,N\pr+N\el}^{\TT}({\cal C}(Q\pr,Q\el))
\end{equation}
where $n_\al(q)$ is the number of loops containing $q$ particles of species~$\al$ in the partition $Q_{\al}$. In~\eqref{A2 Z(C)}, $z_\phi(\LL)$ is a renormalized loop activity
\eq{	\label{def z_phi(LL)}
    z_\phi(\LL) = z(\LL) \e{I_R(\LL)}, \qquad I_R(\LL)=\frac{\be}{2} (V-\phi)(\LL,\LL),
}
and the truncated Mayer coefficient ${\cal B}_{\phi,N}^{\TT}$ is defined by a suitable truncation of the usual Mayer coefficient ${\cal B}_{\phi,N}$ for $N$ loops with pair interactions $\phi$ (see~\cite{b13}). This truncation ensures that ${\cal B}_{\phi,N}^{\TT}$ remains integrable over the relative distances between the loops when $\phi$ is replaced by $V$. The two first truncated Mayer coefficients are ${\cal B}_{\phi,1}^{\TT} = 1$ and
\begin{equation}
{\cal B}_{\phi,2}^{\TT} = \exp (-\beta \phi) - 1 + \beta \phi 
-{\beta^2 \phi^2 \over 2!} + {\beta^3 \phi^3 \over 3!}.
\end{equation}
A vertex corresponding to a cluster $C$ where the associated variables are not integrated over is called a root point (or white point). The integration over an internal (or black) point is performed according to the measure
\begin{equation}
{\cal D}(C) = \sum_{Q\pr,Q\el} \int \prod_{k=1}^{L\pr} \romd \R_k^{(\rmp)} 
\prod_{k=1}^{L\el} \romd \R_k^{(\rme)}
\int \prod_{k=1}^{L\pr} \romD(\X_k^{(\rmp)})
\prod_{k=1}^{L\el} \romD(\X_k^{(\rme)}) 
\end{equation}

\paragraph{Bonds}
Two clusters $C_i$ and $C_j$ are connected by at most one bond ${\cal F}_{\phi}(C_i,C_j)$ which can be either $-\beta \Phi$, $\beta^2 \Phi^2/2!$ or $-\beta^3 \Phi^3/3!$. The potential $\Phi(C_i,C_j)$ is the total interaction potential between the loop clusters  ${\cal C}_i(Q_i^{(\rmp)},Q_i^{(\rme)})$ and ${\cal C}_j(Q_j^{(\rmp)},Q_j^{(\rme)})$ that describe the internal states of $C_i$ and $C_j$ respectively, i.e.
\begin{equation}
\Phi(C_i,C_j) = \Phi(\CC_i,\CC_j) = \sum_{\LL\in\CC_i} \sum_{\LL'\in\CC_j} \phi(\LL,\LL').
\end{equation}

\paragraph{Special rules}	\label{P special rules}
In two cases, the weight~\eqref{A2 Z(C)} of a \emph{black} cluster must be modified to avoid double counting:
\begin{enumerate}
\item[(i)] If $C$ is an intermediate cluster in a convolution $(-\be\Phi)\star(-\be\Phi)$ and contains only a single electron or proton (represented by a loop $\LL$ with $q=1$), then its weight is
\eq{	\label{A2 poids special}
    Z_\phi^{\TT}(C) = z_\phi(\LL) - z(\LL)
}
instead of $Z_\phi^{\TT}(C)=z_\phi(\LL)$.
\item[(ii)] If $C$ is a cluster connected to the rest of the graph by a single bond $\frac{1}{2}(\be\Phi)^2$ and contains only a single electron or proton, then its weight is also given by~\eqref{A2 poids special}.
\end{enumerate}

\paragraph{Diagrammatic expansion of the two-body loop density}
In view of \eqref{A2 chiTk loop}, we need the screened cluster expansion of $\rho_{\TT}(\LL_a,\LL_b)$. According to eq. (4.31) of \cite{b13}, it is given by
\begin{multline}	\label{SCE rho(L_a,L_b)}
\rho_{\text{{\tiny T}}}(\LL_a,\LL_b) = \sideset{}{^*}\sum_G \frac{1}{S_G}\int\romD(C_{ab}) \Big( \sum_{\LL_i,\LL_j \in \CC_{ab}} \delta(\LL_i,\LL_a) \delta(\LL_j,\LL_b) \Big) Z_\phi^{\TT}(C_{ab}) \\
\hspace{3cm} \times \int\prod_k \romD(C_k) Z_\phi^{\TT}(C_k) \Big[\prod\FF_\phi\Big]_G \\
\hspace{1.75cm}+ \sideset{}{^*}\sum_G \frac{1}{S_G}\int\romD(C_a)\romD(C_b) \Big( \sum_{\LL_i \in \CC_a} \delta(\LL_i,\LL_a) \sum_{\LL_k \in \CC_b} \delta(\LL_k,\LL_b) \Big)  \\
\times Z_\phi^{\TT}(C_a) Z_\phi^{\TT}(C_b)  \int\prod_k \romD(C_k) Z_\phi^{\TT}(C_k) \Big[\prod\FF_\phi\Big]_G
\end{multline}
where the graphs $G$ involve either a single root cluster $C_{ab}$ that contains both loops $\LL_a$ and $\LL_b$ (which can be the same loop since coincident points are included), or two roots clusters $C_a$ and $C_b$ with $\LL_a$ in $\CC_a$ and $\LL_b$ in $\CC_b$.
The symmetry factor  $S_{G}$ is the number of permutations of labelled black clusters that leave the product of bonds $[\prod {\cal F}_{\phi}]_{G}$ unchanged (only clusters with identical numbers of protons and electrons are permuted). The sum $\sum_G^*$ runs over all topologically different unlabelled graphs $G$ which are no longer integrable over the relative distances between the clusters $\{C_i\}_{i=1,...,n}$ when $\phi$ is replaced by $V$. Except for this additional constraint (represented by the star in $\sum_G^*$), the graphs $G$ have the same topological structure as the familiar Mayer diagrams: the one-particle points are replaced by particle clusters and the usual Mayer links are now the bonds $\FF_\phi$. A few graphs occuring in the screened cluster expansion of $\chiTk$ are shown on Fig.~1.
%%%%%%%%%%%%%%%%%%%%%%%%%%%%%%%
\begin{figure}[h!]
\begin{center}
    \epsfig{file=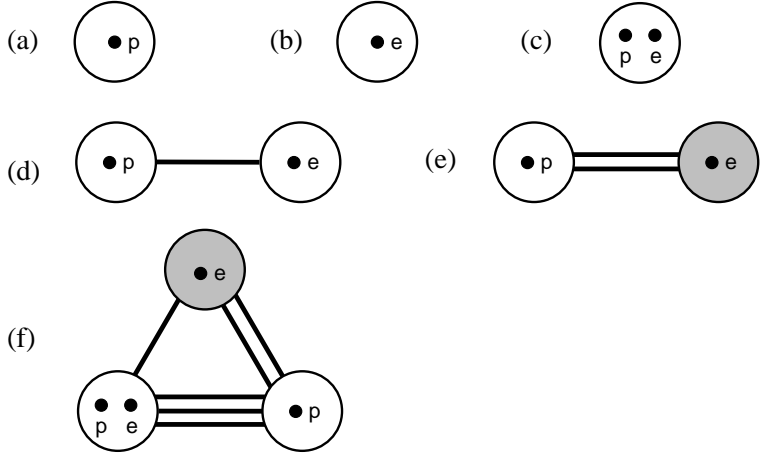}
    \caption{\footnotesize Examples of graphs occuring in the screened cluster expansion of $\chiTk$. The clusters are drawn as circles containing a certain number of protons and electrons. The graphs have either one root (white) cluster $C_{ab}$ (graphs a, b, c and e), or two root clusters $C_a$ and $C_b$ (with $C_a$ drawn to the left of $C_b$). The bonds $\boldsymbol{-\!-}$, $\boldsymbol{=\!=}$ and $\boldsymbol{\equiv\!\equiv}$ are $-\be\Phi$, $(\be\Phi)^2/2$ and $(-\be\Phi)^3/3!$ respectively. 	\label{Fig graphs K}}
\end{center}
\end{figure}

Because of the central role played by the effective potential~$\phi$~\eqref{A2 def graph phi}, it is useful to recall from \cite{BMA} the following exact formula for its Fourier transform
\begin{multline}	\label{A2 TF phi}
\TF{\phi}(\vk,\chia,\chib) = \int\romd\ver\, \exp[-i \vk\cdot\ver]\, \phi(\ver,\chia,\chib) \\
= e_{\al_a}e_{\al_b} \int_0^{q_a}\! \romd s_a \int_0^{q_b}\! \romd s_b \,\e{i\vk\cdot[\lam_a\X_a(s_a) - \lam_b \X_b(s_b)]} \sum_{n=-\infty}^{\infty} \frac{4\pi}{k^2 + \kappa^2(k,n)} \e{-i 2 \pi n (s_a-s_b)}
\end{multline}
where the screening factor for frequency~$n$ is
\begin{equation}
    \kappa^2(k,n) = 4\pi\be\sum_\al e_\al^2 \frac{2z_\al}{(2\pi\lam_\al^2)^{3/2}} \int_0^1 \romd s \int\romD(\vxi)\, \e{i\vk\cdot\lam_\al\vxi(s)} \, \e{i 2 \pi n s}.
 \label{A2 kappa2 int font}
\end{equation}
This is the expression (33) of \cite{BMA} restricted to one-particle loops.
The functional integral can be evaluated using \eqref{A2 cov} with the result
\begin{equation}
    \kappa^2(k,n)= 4\pi\be\sum_\al e_\al^2 \frac{2z_\al}{(2\pi\lam_\al^2)^{3/2}} \int_0^1 \romd s\, \e{-\frac{1}{2}k^2 \lam_\al^2 s(1-s)} \e{i 2 \pi n s}.
\end{equation}
Notice that the zero frequency term $\kappa^2(k,n=0) = \kappa^2(k)$ is identical to~\eqref{A2 def kappa2(k)}.

\section{Dielectric screening in the atomic limit}	\label{S dielectric screening}

The screened cluster expansion allows one to study the response function of the e-p plasma in the atomic limit, without encountering any divergence, thanks to the screening effects embodied in~$\phi$. We show here, by a term by term analysis of the diagrammatic series, that the response function tends in the coupled limit $\be\to\infty$, $k\to 0$ ($\mu\in I_\delta$) defined in~\eqref{A2 THM} to the value $-4\pi\rhoat \alphaH$, as expected for a gas of hydrogen atoms at low density. This result, announced in~\cite{BM_cancun}, constitutes a first principles derivation of a dielectric constant in the quantum e-p plasma.

In the atomic limit, the fugacities $z_\al = \exp[\be\mu_\al]$ (and hence the densities $\rho_\al\ide$) vanish exponentially fast as $\be \to \infty$. The different lengths in the system are therefore ordered according to
\eq{
   \lam_\al \ll \be e^2 \ll (\rhoat)^{-1/3} \ll \kappa^{-1}
}
where the first inequality follows from $\lam_\al \propto \sqrt{\be}$. In the present limit, the screening length $\kappa^{-1}$ diverges and the effective potential $\phi$ tends to the bare Coulomb potential $V$. The difference $\phich \equiv \phi - V$ between $\phi$ and $V$ is therefore expected to be small. It is indeed proven in~\cite{BMA} that at sufficiently small fugacities
\eq{	\label{A2 borne phich}
    |\phich(\ver,\chia,\chib)| \leq K\, q_a q_b e^2 \kappa,
}
where~$K$ is a constant independent of the loop variables. This allows us to replace in all graphs the renormalized loop activities $z_\phi(\LL)$ \eqref{def z_phi(LL)} by the bare activities $z(\LL)$, since
\eq{	\label{A2 dev exp[-be I(LL)]}
    \e{I_R(\LL)} = 1 + \OO(\be e^2 \kappa).
}
Similarly, it will be legitimate at leading order to replace $\phi$ by $V$ in all statistical weights $Z_\phi^{\TT}(C)$.

\subsection{The mean field ionic contribution}	\label{P Mean-field}

We start by calculating the simplest contributions to~$\chiTk$, namely that of free protons and electrons (graphs (a) and (b) of Fig.~1). The statistical weight \eqref{A2 Z(C)} of the root cluster $C_{ab}=C(1,0)$ or $C(0,1)$ involves a single protonic or electronic loop $\LL=(\R,\al,q=1,\vxi)$ with activity $z_\phi(\LL) \to z(\LL) = 2 z_\al / (2\pi\lam_\al^2)^{3/2}$. According to \eqref{A2 chiTk loop} and \eqref{SCE rho(L_a,L_b)}, the contribution to $\chiTk$ of these graphs is
\eq{	\label{A2 K_01 + K_10}
    -\frac{4\pi\be}{k^2} \sum_{\al=\text{e,p}} e_\al^2\frac{2 z_\al}{(2\pi\lam_\al^2)^{3/2}} \int\romD(\vxi) \int_0^1\romd s\,	\e{i\vk\cdot\lam_\al \vxi(s)} = -\frac{\kappa^2(k)}{k^2}.
}
Eq. \eqref{A2 K_01 + K_10} is identical to the first term \eqref{A2 -kappa^2(k)/k^2} of the ``na\"{\i}ve'' virial expansion. The mean field (or Debye-H\"{u}ckel) result~\eqref{A2 D-H} is obtained by adding to~\eqref{A2 K_01 + K_10} the contributions of the four graphs
\eq{	\label{4 o-o}
    \epsfig{file=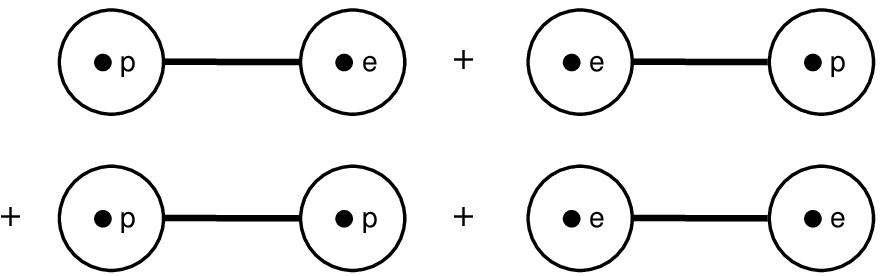}
}
At leading order, \eqref{4 o-o} becomes
\begin{equation}	\label{A2 fat}
    -\frac{4\pi\be}{k^2} \sum_{\substack{\al_a=\text{e,p}\\\al_b=\text{e,p}}}
    e_{\al_a} e_{\al_b} \int\romD(\vxi_a)\romD(\vxi_b)\int_0^1\romd \tau
    \e{i\vk\cdot\lam_b\vxi_b(\tau)} z(\LL_a) z(\LL_b)	
     (-\be\TF{\phi}(\vk,\chi_a,\chi_b)).
\end{equation}
Using~\eqref{A2 TF phi}, \eqref{A2 def kappa2(k)} and
\eq{	\label{super prop}
    \int_0^1\romd \tau \int_0^1\romd s\, \e{i 2 \pi n s} \int \romD(\vxi)\, z(\LL)
    \e{i\vk\cdot\lam[\vxi(\tau)-\vxi(s)]} = \delta_{n,0} \int_0^1\romd \tau\int\romD(\vxi)\,
    	z(\LL) \e{i\vk\cdot\lam\vxi(\tau)},
}
which follows from the periodicity of $\vxi(s)$ and of $\exp[i 2 \pi n \tau]$, the contribution~\eqref{A2 fat} is easily evaluated, and gives, when added to~\eqref{A2 K_01 + K_10},
\eq{	\label{MF result}
    \chiTMFk \equiv -\frac{\kappa^2(k)}{k^2} + \frac{\kappa^4(k)}{k^2[k^2+\kappa^2(k)]} = -\frac{\kappa^2(k)}{k^2 + \kappa^2(k)}.
}
This result agrees with the Debye-H\"{u}ckel formula~\eqref{A2 D-H} since $\kappa^2(k) \to \kappa^2$ in the limit \eqref{A2 THM}. Notice that \eqref{MF result} saturates the perfect screening relation~\eqref{A2 -1}. The sum of all other contributions to $\chiTk$ must therefore vanish at $k=0$.

\subsection{The atomic contribution}	\label{P atomic contribution}

\noi The dielectric screening effect~\eqref{A2 chi_ep} is expected to be contained in the graph consisting in a single root cluster $C(1,1)$ (graph (c) of Fig.~1), since the latter graph describes an interacting electron-proton pair (H atom). We calculate here the contribution of this graph, with a precise mathematical control on its asymptotic value in the atomic limit. The cluster~$C(1,1)$ is made of one protonic loop $\LL\pr = (\ver\pr,\text{p},1,\vxi\pr)$ and one electronic loop $\LL\el = (\ver\el,\text{e},1,\vxi\el)$. Its contribution reads
\begin{multline}	\label{A2 K_11}
    -\frac{4\pi\be}{k^2} \int\romd\ver\el\int\romd\ver\pr\int\romD(\vxi\el)\int\romD(\vxi\pr)
    \sum_{\substack{i=\text{e,p}\\j=\text{e,p}}} e_i e_j \int_0^1\romd s\,
    	\e{i\vk\cdot\lam_j\vxi_j(s)} \e{-i\vk\cdot\ver_i} \delta(\ver_j) \\ \times z_\phi(\LL\el)z_\phi(\LL\pr)
    	{\cal B}^{\TT}_{\phi,2}(\LL\el,\LL\pr).
\end{multline}
Using the translation invariance ${\cal B}^{\TT}_{\phi,2}(\LL\el,\LL\pr) = {\cal B}^{\TT}_{\phi,2}(\ver\el-\ver\pr,\chi\el,\chi\pr)$ and introducing the change of variables $\ver\pr \tend -\ver$, we can rewrite~\eqref{A2 K_11} as
\begin{multline}	\label{A2 98}
    -\frac{4\pi\be e^2}{k^2} \int\romd\ver \int\romD(\vxi\el)\int\romD(\vxi\pr)\int_0^1\romd s
    	\frac{2z\el}{(2\pi\lam\el^2)^{3/2}}\frac{2z\pr}{(2\pi\lam\pr^2)^{3/2}}
    	\e{I_R(\LL\el)} \e{I_R(\LL\pr)} \\
    	 \Big(
    	\e{i\vk\cdot\lam\el\vxi\el(s)} + \e{i\vk\cdot\lam\pr\vxi\pr(s)}
    	-\e{-i\vk\cdot\ver} \e{i\vk\cdot\lam\pr\vxi\pr(s)}
    	-\e{i\vk\cdot\ver}  \e{i\vk\cdot\lam\el\vxi\el(s)}
    	\Big)
    	 {\cal B}^{\TT}_{\phi,2}(\ver,\chi\el,\chi\pr).
\end{multline}
In the atomic limit, the factors~$\exp[-\be I_R(\LL)]$ go to~1, and the effective potential~$\phi$, which enters the Mayer coefficient~${\cal B}^{\TT}_{\phi,2}$, tends to the bare Coulomb potential~$V(\ver,\chia,\chib)$. Recall that the truncation in ${\cal B}^{\TT}_{\phi,2}$ is such that ${\cal B}^{\TT}_{\phi,2}$ remains integrable when $\phi$ is replaced by~$V$. The integral in~\eqref{A2 98} is hence finite for any~$\vk$, but since it is divided by $k^2$ and $k \to 0$ in the dielectric limit defined in~\eqref{A2 THM}, it is important to note that the integral behaves at small $k$ as $k^2$ times
\eq{
   \int\romd\ver\, \int\romD(\vxi\el)\int\romD(\vxi\pr) (\hat{\vk}\cdot e \ver)\cdot (\hat{\vk}\cdot e [\ver+\lam\el\vxi\el(s) - \lam\pr \vxi\pr(s)]) {\cal B}_{\phi,2}^{\TT}(\ver,\chi\el,\chi\pr).
}
Since the integrand in the above expression becomes non integrable at zero density (it decays as $1/r^2$ when $\phi$ is replaced by $V$), we further truncate ${\cal B}^{\TT}_{\phi,2}$ by introducing the decomposition
\eq{	\label{A2 def W_V^TT}
    {\cal B}^{\TT}_{\phi,2}(\LL_a,\LL_b) = {\cal B}^{\TT\TT}_{\phi,2}(\LL_a,\LL_b)
    + \frac{1}{4!}\parenth{-\be\phi(\LL_a,\LL_b)}^4
    + \frac{1}{5!}\parenth{-\be\phi(\LL_a,\LL_b)}^5,
}
which defines ${\cal B}^{\TT\TT}_{\phi,2}(\LL_a,\LL_b)$. Now ${\cal B}^{\TT\TT}_{\phi,2}(\LL_a,\LL_b)$ decays as $|\R_a-\R_b|^{-6}$, so that second moments are finite. The terms $\phi^4$ and $\phi^5$ give a negligible contribution in the atomic limit, as compared to the free charges contribution (see section~6.3).
After a little algebra, the difference between ${\cal B}^{\TT\TT}_{\phi,2}$ and ${\cal B}^{\TT\TT}_{V,2}$ is found to be
\eq{	\label{A2 132}
    {\cal B}^{\TT\TT}_{\phi,2}(\LL_a,\LL_b) = {\cal B}^{\TT\TT}_{V,2}(\LL_a,\LL_b) + R_1 + R_2 + R_3,
}
where the remaining terms $R_i$ are
\begin{gather}
\label{A2 def R_1}
    R_1(\LL_a,\LL_b) = (\e{-\be\phich(\LL_a,\LL_b)}-1) {\cal B}^{\TT\TT}_{V,2}(\LL_a,\LL_b) \\
\label{A2 def R_2}
    R_2(\LL_a,\LL_b) = \sum_{\substack{n,m=0 \\ n+m\geq 6}}^5 \frac{1}{n!} \frac{1}{m!}
    	(-\be\phich(\LL_a,\LL_b))^n (-\be V(\LL_a,\LL_b))^m \\
\label{A2 def R_3}
    R_3(\LL_a,\LL_b) = \sum_{n=6}^\infty \frac{1}{n!} 
    	(-\be\phich(\LL_a,\LL_b))^n \sum_{m=0}^5 (-\be V(\LL_a,\LL_b))^m.
\end{gather}
We define $\chiTatk$ as \eqref{A2 98} with bare activities and with ${\cal B}^{T}_{\phi,2}$ replaced by ${\cal B}^{\TT\TT}_{V,2}$:
\begin{multline}	\label{A2 def chi_at}
    \chiTatk \equiv -\frac{4\pi\be e^2}{k^2} \int\romd\ver  \int\romD(\vxi\el)\int\romD(\vxi\pr)\int_0^1\romd s
    	\frac{2 z\el}{(2\pi\lam\el^2)^{3/2}}\frac{2 z\pr}{(2\pi\lam\pr^2)^{3/2}}
    	 \, {\cal B}^{\TT\TT}_{V,2}(\ver,\chi\el,\chi\pr) \\ \times \Big(
    	\e{i\vk\cdot\lam\el\vxi\el(s)} + \e{i\vk\cdot\lam\pr\vxi\pr(s)}
        	-\e{-i\vk\cdot\ver} \e{i\vk\cdot\lam\pr\vxi\pr(s)}
    	-\e{i\vk\cdot\ver}  \e{i\vk\cdot\lam\el\vxi\el(s)} \Big) .
\end{multline}
To evaluate \eqref{A2 98} in the atomic limit, we must establish the following two points:
\begin{itemize}
\item[(i)] \emph{Replacing $\phi$ by $V$}. We must show that \eqref{A2 def chi_at} is indeed the leading behaviour of \eqref{A2 98} in the atomic limit. This is done in appendix (section A.1), where we prove that the contribution of the remaining terms $R_i$ in \eqref{A2 132} are $o(\rhoat)$, and hence negligible as $\be \to \infty$.
\item[(ii)] \emph{Excited and ionized states}. We must show that the contributions to~$\chiTk$ of excited and ionized states of the hydrogen atom, which are included in \eqref{A2 def chi_at}, are also negligible in the atomic limit. This is done in section A.2 of the appendix.
\end{itemize}

Using point (i), we proceed to the evaluation of~$\TF{\chi}_{\text{at}}(\vk)$ at leading order. We introduce the relative and center of mass coordinates, defined by
\eq{	\label{A2 Coord cm/rel}
    \begin{cases}
    \lam\el\vxi\el(s) = \lamat\vxi\atH(s) + \frac{m\pr}{M}\lam\vxi(s) \\
    \lam\pr\vxi\pr(s) = \lamat\vxi\atH(s) - \frac{m\el}{M}\lam\vxi(s)
    \end{cases}
}
with
\eq{
   \lam=\hbar\sqrt{\frac{\be}{m}}, \quad m = \frac{m\el m\pr}{M}, \quad M=m\el + m\pr
}
Notice that $\lam\el\lam\pr = \lamat\lam$. It is easy to verify that the Gaussian measure $\romD(\vxi\el)\romD(\vxi\pr)$ and $\romD(\vxi\atH)\romD(\vxi)$ have the same covariance. In these new variables, the functional integrations over $\vxi\atH$ and $\vxi$ factorise. The center of mass integration
\eq{
    \frac{4}{(2\pi\lamat^2)^{3/2}} \int\romD(\vxi\atH)\e{i\vk\cdot\lamat\vxi\atH(s)} = f_s(\vk)
}
gives the factor~$f_s(\vk)$ already encountered in~\eqref{A2 f_tau(k)}. $\TF{\chi}_{\text{at}}(\vk)$ hence becomes
\begin{multline}	\label{A2 111}
    \TF{\chi}_{\text{at}}(\vk) = -\frac{4\pi\be e^2}{k^2} 4\e{2\be\mu} \int_0^1\romd s  \, f_s(\vk)
    	\int\romd\ver\int\romD(\vxi)\, {\cal B}_{V,2}^{\TT\TT}(\ver,\lam\vxi) \\
    	\Big(
    	\e{i\vk\cdot\frac{m\pr}{M}\lam\vxi(s)} + \e{-i\vk\cdot\frac{m\el}{M}\lam\vxi(s)}
    	-\e{-i\vk\cdot\ver} \e{-i\vk\cdot\frac{m\el}{M}\lam\vxi(s)}
    	-\e{i\vk\cdot\ver}  \e{i\vk\cdot\frac{m\pr}{M}\lam\vxi(s)}
    	\Big)
\end{multline}
where, according to~\eqref{A2 def W_V^TT}, we have
\begin{equation}	\label{A2 pos W_V^TT}
    {\cal B}_{V,2}^{\TT\TT}(\ver,\lam\vxi) = \exp\Big[\be e^2\! \int_0^1\romd s V(\ver+\lam\vxi(s))\Big] 
    - \sum_{n=0}^5 \frac{1}{n!}\parenth{\be e^2\! \int_0^1\romd s V(\ver+\lam\vxi(s))}^n >0.
\end{equation}
In order to determine to low temperature limit of~\eqref{A2 111} in terms of atomic eigenvalues and eigenstates, we convert this expression back into operator's language. Notice that the factor in brackets in~\eqref{A2 111} can be rewritten as
\eq{
    \parenth{\e{i\vk\cdot\frac{m\pr}{M}[\ver+\lam\vxi(s)]} - \e{-i\vk\cdot\frac{m\el}{M}[\ver+\lam\vxi(s)]}} \parenth{\e{-i\vk\cdot\frac{m\pr}{M}\ver} - \e{i\vk\cdot\frac{m\el}{M}\ver}}.
}
Defining the operator $\opA = \e{i\vk\cdot\frac{m\el}{M}\opq} -\e{-i\vk\cdot\frac{m\pr}{M}\opq}$ as in~\eqref{A2 def A_nn'}, we find from the Feynman-Kac formula~\cite{SimSchRoe} that~\eqref{A2 111} is equivalent to
\renewcommand{\Ao}{\opA}
\renewcommand{\Vo}{\opV}
\begin{multline}	\label{A2 OKOK}
    \TF{\chi}_{\text{at}}(\vk) = -\frac{4\pi e^2}{k^2} 4 \e{2\be\mu}\int_0^\be\romd \tau    f_{\tau/\be}(\vk) \,
    \Tr \Big\{\opU(\be-\tau)\opA\adj\opU(\tau)\opA \\
    -\e{-\be\opH_0}\opT \Big[ \Ao\adj(\tau)\opA
    \sum_{n=0}^5 \frac{1}{n!} \Big( \int_0^\be\romd s  \overline{\opV}(s) \Big)^n \Big] \Big\}.
\end{multline}
In~\eqref{A2 OKOK}, the trace runs over the spectrum of the hydrogen Hamiltonian
\eq{
    \opH = \opH_0 + \opV, \qquad \opH_0 = \frac{\vp^2}{2m}, \qquad \opV = -\frac{e^2}{|\opq|}, \qquad \overline{\opV} = -\opV,
}
$\opU(s)=\exp[-s \opH]$ is the evolution operator, and the freely time-evolved operators are defined by
\eq{
    \Ao(s) = \e{s\opH_0} \opA \e{-s\opH_0}, \qquad
    \Vo(s) = \e{s\opH_0} \opV \e{-s\opH_0}.
}
The subtraction of the freely evolving quantities in~\eqref{A2 OKOK} ensures the finiteness of the trace. The dominant low temperature terms will come from the ground state contribution of $\opU(s)$ when evaluating the trace. Let $\op{P}=\ket{0}\bra{0}$ be the projector on the ground state, $\opQ = \id -\op{P}=\sum_{\m\geq 1}\ket{\m}\bra{\m}$, and decompose $\opU(s) = \opU_{\opP}(s)+\opU_{\opQ}(s)$ accordingly, i.e. $\opU_{\opP}(s)=\exp[-E_0 s]\opP$ and $\opU_{\opQ}(s)=\opU(s)Q = Q\opU(s)$. We split~\eqref{A2 OKOK} into
\eq{	\label{A2 split chi_at = 0+1}
     \TF{\chi}_{\text{at}}(\vk) = \TF{\chi}_{\text{at}}^{(0)}(\vk) +  \TF{\chi}_{\text{at}}^{(1)}(\vk),
}
where
\begin{multline}	\label{A2 Ground State}
   \TF{\chi}_{\text{at}}^{(0)}(\vk) = -\frac{4\pi e^2}{k^2} 4 \e{2\be\mu}\int_0^\be\romd \tau    f_{\tau/\be}(\vk) \,
    \Tr \Big\{ \opU_{\opP}(\be-\tau) \opA\adj\opU_{\opP}(\tau)\opA \\
    +\opU_{\opQ}(\be-\tau)\opA\adj\opU_{\opP}(\tau)\opA	
    +\opU_{\opP}(\be-\tau)\opA\adj\opU_{\opQ}(\tau)\opA \Big\}
\end{multline}
is the part of~$\TF{\chi}_{\text{at}}(\vk)$ which has contributions from the ground state, and $\TF{\chi}_{\text{at}}^{(1)}(\vk)$ involves only contributions from excited and ionized states of the hydrogen atom, as well as the truncations terms in~\eqref{A2 OKOK}. It is shown in appendix (section A.2) that the part $\TF{\chi}_{\text{at}}^{(1)}(\vk)$ is negligible in the atomic limit as compared to $\TF{\chi}_{\text{at}}^{(0)}(\vk)$. In terms of atomic eigenvalues and eigenstates, \eqref{A2 Ground State} becomes
\eq{	\label{def chi_at^0}
       \TF{\chi}_{\text{at}}^{(0)}(\vk) = -\frac{4\pi e^2}{k^2} 4 \e{2\be\mu}
       \int_0^\be\romd \tau    f_{\tau/\be}(\vk) \,
       \sum_{\substack{\n,\n'\geq 0\\\n \cdot \n' = 0}}
       \e{-(\be-\tau)E_{\n}} \e{-\tau E_{\n'}} |A_{\n\n'}(\vk)|^2
}
where $A_{\n\n'}(\vk)$ is defined in~\eqref{A2 def A_nn'}. Factoring out the atomic density~\eqref{A2 def rho_at}, we find ($E_0=E\atH$)
\eq{	\label{A2 106}
     \TF{\chi}_{\text{at}}^{(0)}(\vk) = -4\pi\rhoat\, \alphaH(\vk,\beta),
}
where $\alphaH(\vk,\beta)\simeq\alphaH$ if $k\ll\lam\atH^{-1}$ (see~\eqref{A2 formule alpha}). The graph~$C_{ab}=C(1,1)$ does therefore indeed describe, at leading order in the atomic limit, the dielectric screening due to the polarisation of the hydrogen atoms, in agreement with the anticipated result~\eqref{A2 chi_ep}. We stress that no divergences occur in the present derivation of \eqref{A2 106}, in contradistinction to the elementary calculation of section~\ref{S elementary}.

\subsection{Higher order contributions}	\label{SS other graphs}

We consider all graphs in the screened cluster expansion different from the pure atomic graph $C_{ab}=C(1,1)$, and argue that they give higher order contributions to $\chiTk$ in the atomic limit $\be\to\infty$, $\mu\in I_\delta$, if $k \gg \lam_{\rm I}^{-1}$ (see~\eqref{A2 THM}). In fact, the estimations of those graphs differ from the estimations presented in~\cite{8bis} for the particles densities $\rho_\al =\sum_q q \int\romD(\X)\rho(\LL)$ only by the different weight attached to the root points. The definitions of weights for black clusters and bonds are indeed the same, so that the factors arising from integrations over these clusters can be evaluated in the same way. In \cite{8bis}, an analysis of the behaviour of Coulomb multiparticle Green functions shows that these factors vanish exponentially fast as $\be \to \infty$. We do not reproduce this mathematical analysis here, but merely adapt the arguments to the present situation.
%We show below on simple examples how the order of magnitude of the graphs
%can be estimated.

We assume without loss of generality that the set of fugacities $\{z\el,z\pr\}$ satisfy the pseudo-neutrality condition
\eq{	\label{A2 pseudo}
   \sum_{\al = \text{e,p}} e_\al \frac{2 z_\al}{(2\pi\lam_\al^2)^{3/2}} = 0,
}
so that electrical neutrality $\rho\el\ide=\rho\pr\ide$ of the ideal system (no Coulomb interactions) holds. With this choice, the difference between the chemical potentials $\mu\el$ and $\mu\pr$ are given by \eqref{A2 2.5}, and $\chiTk$ depends on the fugacity
\eq{
    z = \e{\be \mu}, \qquad \mu = \frac{\mu\el+\mu\pr}{2}.
}
The choice of fugacities satisfying the pseudo-neutrality condition will allow to greatly reduce the number of graphs contributing at a given order. Moreover, we choose $\mu>E\atH$ in the interval $I_\delta$ so that, according to~\eqref{Strong inequ}, collective screening effects are due to ionized electrons and protons with densities $\rho\el\ide=\rho\pr\ide$ (see end of section~\ref{A2 S Atomic limit}). The estimations will be performed in terms of the exponentially small factor $\rho\el\ide\propto z=\exp[\be\mu]$, $\mu\in I_\delta$, disregarding any power law dependence in $\be$. The condition \eqref{Strong inequ} together with $k \gg \lami^{-1}$, or equivalently
\eq{	\label{neglige ionic}
   \frac{\be e^2 \rho\ide_{N\pr N\el}}{k^2} \ll \frac{\be e^2 \rho\el\ide}{k^2}  \ll \aB^3 \rho\atH\ide,
}
will allow simple estimations of the graphs that avoid considering their small $k$ behaviour.

\paragraph{(a) The graph consists in the single root cluster $C(N\pr,N\el)$} \hfill \phantom{x}\\
\indent\: We assume $(N\pr,N\el)\neq(1,0),(0,1),(1,1)$, since these cases have already been considered previously. The contribution of the root cluster is
\begin{align}	\notag
\psfrag{NN}{$N\pr,N\el$}
   \raisebox{1mm}{\Graph{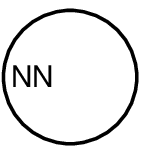}}\;		%figCpe.eps 
   &=  -\frac{4\pi\be}{k^2} \int\romD(C) \sum_{\LL_i,\LL_j \in C} e_{\al_i} q_i e_{\al_j} \int_0^{q_j}\romd \tau \e{-i\vk\cdot\R_i} \e{i\vk\cdot\lam_j\X_j(\tau)}\delta(\R_j) Z_\phi^{\TT}(C) \\ \label{def g}
   &=   \frac{\be e^2}{k^2} \, g_{N\pr N\el}(k),
\end{align}
At leading order, the effective potential $\phi$ in the renormalized activities $z_\phi$ and in the Mayer coefficients ${\cal B}_{\phi,N}^{\TT}$ can be replaced by the bare potential~$V$. The loop integrals in $g_{N\pr N\el}(k)$ do indeed converge at zero density (for any $k>0$), despite the long range of the Coulomb interaction, because of the truncation built in ${\cal B}_V^{\TT}$. With these replacements, all density dependences in $g_{N\pr N\el}(k)$ are contained in the prefactors of the integrals, which are obviously of order $z^{N\pr + N\el}$. To evaluate the low temperature behaviour of the integrals, we convert them back in operator's language using the Feynman-Kac formula. Similarly to~\eqref{A2 OKOK}, $g_{N\pr N\el}(k)$ is given by a trace evaluated over suitably antisymmetrized states of the $N\pr$ protons and $N\el$ electrons (because of the sum over the partitions), of a time ordered product of the Gibbs operator $\exp[-\be \opH_{N\pr N\el}]$ and operators $\exp[i\vk\cdot\opq_i(s)]$, where $\opq_i(s)$ is the time evolved position operator for the $i^{\text{th}}$ particle. Notice that the truncation terms ensure the finiteness of this trace despite the long range of the Coulomb interaction. A typical term in the truncation involves Gibbs operators for sub-clusters of $(M\pr,M\el)\neq(N\pr,N\el)$ particles, $M\pr\leq N\pr$, $M\el \leq N\el$. As $\be \to \infty$, the leading behaviour of the truncated trace is controlled by the ground state contribution of $\exp[-\be \opH_{N\pr N\el}]$, which is proportional to $\exp[-\be E_{N\pr N\el}]$ (discarding powers of~$\be$). As we have shown in details in the case of the atomic contribution (see appendix~A), the excited states and the truncation terms are expected to give exponentially smaller contributions at low temperatures. The function $g_{N\pr N\el}(k)$ is hence expected to behave as
\eq{	\label{estim g}
    g_{N\pr N\el}(k) \propto \e{-\be(E_{N\pr N\el} -\mu(N\pr + N\el))}, \qquad \be \to \infty,
}
where $E_{N\pr N\el}$ is the infimum of the spectrum of $\opH_{N\pr N\el}$. Using~\eqref{Strong inequ}, $g_{N\pr N\el}(k)$ is therefore bounded for $\be$ large by an expression exponentially smaller than the density of ionized charges:
\eq{	\label{borne g}
   g_{N\pr N\el}(k) \leq \rho\el\ide \e{-\be \Gamma},
   \qquad (N\pr,N\el)\neq(1,0),(0,1),(1,1),
}
where $\Gamma$ is a positive constant. In~\cite{b13}, the low temperature behaviour of the truncated traces of Gibbs operators $\exp[-\be \opH_{N\pr N\el}]$ are studied in details, and shown to satisfy the upper bound~\eqref{borne g}. This bound, combined with \eqref{neglige ionic}, is sufficient to show that the graphs consisting of a single root cluster $C(N\pr,N\el) \neq C(1,1)$ do not contribute to the response function in the dielectric regime characterized by the limit~\eqref{A2 THM}.

\paragraph{(b) A black cluster $C_{10}$ or $C_{01}$ is connected by a bond $-\be\Phi$ to a root cluster $C(N\pr,N\el)$}
\eq{	\label{Graph ()---(C_01)}
\psfrag{NN}{$N\pr,N\el$}
\psfrag{+}{$+$}
    \raisebox{1mm}{\Graph{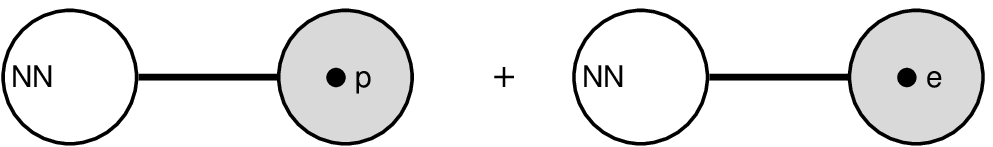}}
}
The integration over the black cluster involves the expression
\eq{	\label{Int lien Phi}
    \sum_{\al=\text{e,p}} \int\romd\R\int\romD(\X)\, z(\LL)\e{I_R(\LL)}
     (-\be\phi(\LL,\LL_i))
}
where $\LL=(\al,q=1,\R,\X)$ is the loop in the black cluster and $\LL_i$ is one of the loops in the cluster $\CC_{N\pr N\el}$. Eq.~\eqref{Int lien Phi} can be evaluated by using the Fourier transform $\TF{\phi}(\vp,\chi,\chi_i)$ \eqref{A2 TF phi} with wave-numbers $|\vp| \to 0$. Using the low $p$ behaviour $\kappa^2(p,n)=\kappa^2 \delta_{n,0} + \gamma_n p^2 + \OO(p^4)$ and the rotational invariance of $z\exp[I_R(\LL)]$, we find that the contributions of the terms $n \neq 0$ vanish by parity in the limit $p \to 0$:
\begin{multline}	\label{5.192}
  \lim_{p \to 0} \int\romD(\X) z\,\e{I_R(\chi)} \sum_{n \neq 0} \int_0^q \romd s\,
  \int_0^{q_i}\romd s'\, (\e{i\vp\cdot\lam_\al\X(s)}-1) (\e{i\vp\cdot\lam_{\al_i}\X_i(s')}-1) \\ \times
  \frac{4\pi}{p^2 (1+\gamma_n)} \e{-i 2 \pi n (s-s')} = 0.
\end{multline}
In~\eqref{5.192}, the two subtractions $-1$ could be freely introduced because $n \neq 0$. Only the term $n=0$ does therefore contribute to~\eqref{Int lien Phi}, and we find
\eq{	\label{1/kappa2}
   \eqref{Int lien Phi} = -\frac{\be e_{\al_i} q_i}{\kappa^2} \sum_{\al=\text{e,p}} e_\al z_\al \frac{2}{(2\pi\lam_\al^2)^{3/2}} \int\romD(\X) \e{I_R(\LL)} = \OO(\be e^2 \kappa)
}
The estimate $\OO(\be e^2 \kappa)$ is obtained by using \eqref{A2 def kappa2} and \eqref{A2 dev exp[-be I(LL)]}, and noting that the term of order~1 vanishes because of the pseudo-neutrality condition~\eqref{A2 pseudo}. The two graphs \eqref{Graph ()---(C_01)} give hence, according to~\eqref{1/kappa2}, a contribution exponentially smaller than the one associated to the graph \eqref{def g}. Notice that without the choice of pseudo-neutrality, \eqref{1/kappa2} would be of order one, and ``dressing'' points in a graph with a black cluster $C_{10}$ or $C_{01}$ connected by a bond $-\be\Phi$ would not raise the order in density.

The leading behaviour \eqref{1/kappa2} can be obtained more directly by replacing in \eqref{Int lien Phi} the effective loop potential $\phi(\LL_1,\LL_2)$ by the Debye potential
\eq{	\label{Debye boucles}
    q_1 q_2 \frac{\exp[-\kappa |\ver|]}{|\ver|}, \qquad \ver=\R_1-\R_2 .
}
The result \eqref{1/kappa2} follows then from the integral $\int\romd\ver \exp[-\kappa r]/r = 4\pi/\kappa^2$. The replacement of $\phi$ by \eqref{Debye boucles} is a priori valid only for distances $r\sim\kappa^{-1}$ (see section~\ref{S loop representation}), but it can be used at all distances to evaluate the leading behaviour of~\eqref{Int lien Phi} (see~\cite{b13}). Indeed, at short distances $r< \lam$, $\phi(\LL_1,\LL_2)$ is given at lowest order by the bare Coulomb potential $V(\LL_1,\LL_2)$ (see~\eqref{A2 borne phich}). The contribution of the region $r< \lam$ to the integral \eqref{Int lien Phi} is of order $\lam^2$, as can be seen after the change of variables $\ver=\lam\vx$. Since $\lam \ll \kappa^{-1}$, this contribution is small as compared to the contribution of the region $\lam<r<\kappa^{-1}$, which is of order $\kappa^{-2}$. At very large distances, $r\gg \kappa^{-1}$, $\phi$ decays as a dipolar potential. This term vanishes however by parity after integration over~$\X$, and the following term in the multipolar expansion is proportional to $\lam^3/r^4$. This term is integrable and the contributions to the integral~\eqref{Int lien Phi} of the large distances $r\gg \kappa^{-1}$ are hence also negligible (they are of order $\kappa\lam^3$), as the short-range contributions. It is therefore legitimate to replace $\phi$ by~\eqref{Debye boucles} at all distances, since we correctly capture in this way the leading contributions to the integral, which come from the intermediate regions $r\sim\kappa^{-1}$. We stress that the exponentially growing factor $1/\kappa^2$ in the estimate~\eqref{1/kappa2} has its origin in the fact that the bond $\phi$ becomes non integrable as the density goes to zero in the atomic limit ($\phi \to V$). Because of this density dependence of the bonds in the graphs, the lowest order at which a graph contributes cannot be deduced immediately from the number of black clusters it contains.

\paragraph{(c) A black cluster $C(1,1)$ is connected by a bond $-\be\Phi$ to the root cluster $C(N\pr,N\el)$:}
\eq{	\label{Graph C-Phi-C primo}
\psfrag{NN}{$N\pr,N\el$}
    \Graph{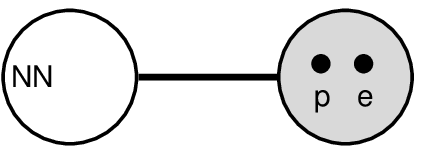}	%Fig_LienPhiCpe
}
For an electron-proton pair forming a hydrogen atom \eqref{Debye boucles} should be replaced by
\eq{	\label{195 ter}
   -e^2 \Big[ \frac{\e{-\kappa |\ver|}}{|\ver|} - \frac{\e{-\kappa |\ver+\vect{a}|}}{|\ver+\vect{a}|} \Big]
}
where $\vect{a}$ is of the order of the Bohr radius. The bond \eqref{195 ter} remains integrable as $\kappa \to 0$ due to neutrality and rotational invariance (the average dipole moment $e\vect{a}$ of the atom is zero). The graph gives hence a contribution to $\chiTk$ that is exponentially smaller than \eqref{def g} because of the additional factor $\rho\atH\ide$ arising from the weight of the black cluster.
%As a consequence, the $\ver$-integral of~\eqref{195 ter} is $\OO(\kappa^{-1})=\OO((\rho\el\ide)^{-1/2})$ as $\kappa \to 0$. Integration on the black cluster gives thus a factor $\rho\atH\ide (\rho\el\ide)^{-1/2}\propto \exp\crochet{-\be(E\atH-\frac{3}{2}\mu)}$ which is exponentially small if $\mu \in ]E\atH,E\atH+\delta]$ is chosen sufficiently close to~$E\atH$. The whole contribution to $\chiTk$ is thus exponentially smaller than that of~\eqref{def g}. Neutrality of the $\CC_{11}$ cluster plays again a crucial role.

\paragraph{(d) A black cluster $C(M\pr,M\el)$ is connected by a bond $-\be\Phi$ to a root cluster $C(N\pr,N\el)$:}
\eq{	\label{Graph C-Phi-C}
\psfrag{NN}{$N\pr,N\el$}
\psfrag{MM}{$M\pr,M\el$}
    \Graph{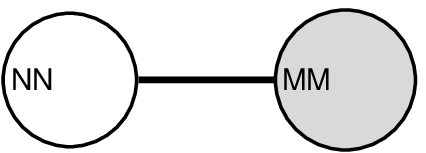}		%Fig_LienPhiCpe
}
The leading contributions associated to the integral over the center of mass of the black cluster arise as before from inter-cluster distances of the order of $\kappa^{-1}$, where the loop potential can be replaced by~\eqref{Debye boucles}. At these distances, the bond $-\be\Phi$ is insensitive to the precise form of the loops within the clusters, since $\lam_\al \ll \kappa^{-1}$. The integration over the center of mass provides as before a factor $\be e^2/\kappa^{2}$, while the integrations over the relative distances in the cluster yield at low temperatures a factor $\rho_{N\pr N\el}\ide$, which arise from the ground state contribution in the associated truncated trace. This provides the estimate $\be e^2 \rho_{N\pr N\el}\ide/\kappa^2$ for the integration over the black cluster in~\eqref{Graph C-Phi-C}. Using \eqref{Strong inequ}, we conclude that the integration over the black cluster $\CC_{N\pr N\el}$ gives, if $(M\pr,M\el)\neq (0,1),(1,0),(1,1)$, a factor vanishing exponentially fast as $\be\to\infty$. Notice that this estimate is also valid, but not optimal, in the case of a neutral cluster (see point~(c) above).

\paragraph{(e) The cases (b), (c) and (d) when the bond is $\frac{1}{2}(\be\Phi)^2$ or $\frac{1}{6}(-\be\Phi)^3$, for instance:}
\eq{	\label{Graphs Phi2 Phi3}
\psfrag{NN}{$N\pr,N\el$}
\psfrag{MM}{$M\pr,M\el$}
    \Graph{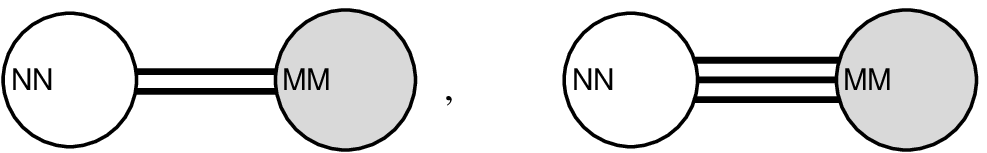}	%Fig_LienPhi23
}
We can argue that these graphs give smaller contributions than those discussed in (b), (c) and (d). This is due to the fact that these bonds give less divergent integrals as $\kappa \to 0$ than the bond $-\be\Phi$. The integration of a bond $\frac{1}{2}(\be\Phi)^2$ provides for example a factor proportional to $\kappa^{-1}\propto z^{-1/2}$. This can be seen by replacing as before the effective potential by the Debye potential \eqref{Debye boucles}, and evaluating
\eq{
    \int\romd\ver\, \frac{\e{-2\kappa r}}{r^2} = \frac{2\pi}{\kappa}.
}
In the case of the bond $\frac{1}{6}(-\be\Phi)^3$, we have to use the correct behaviour of $\phi$ at short distances ($r<\lam$) to estimate the integral of the bond, because the Debye form \eqref{Debye boucles} would lead to logarithmic divergences. As shown in~\cite{b13}, the result is that the integral grows like $\ln(\kappa \lam)$ as $\be \to \infty$. The factor arising from the integration over a cluster connected by a bond $\frac{1}{2}(\be\Phi)^2$ or $\frac{1}{6}(-\be\Phi)^3$ are therefore smaller than the corresponding factor for a bond $-\be\Phi$, and the graphs considered in (b), (c) and (d) are negligible in the atomic limit. With the decomposition \eqref{A2 def W_V^TT} of $\BB_{\phi,2}^{\TT}$, we introduced two graphs with bonds $\frac{1}{4!}(-\be\phi)^4$ or $\frac{1}{5!}(-\be \phi)^5$ between an electron and a proton. Since these bonds are integrable at zero density\footnote{They decrease as $r^{-4}$ and $r^{-5}$ at large distances and the Coulomb singularity at the origin is smoothed out by the functional integration --- see \eqref{A2 smooth}.}, such graphs are of order $z^2/k^2$ (discarding powers of $\be$), and are hence negligible in the dielectric limit \eqref{A2 THM}, in view of \eqref{neglige ionic}.

The above analysis can be generalized to estimate the order of magnitude of an arbitrary graph with one root cluster and black clusters connected by bonds $-\be\Phi$,  $\frac{1}{2}(\be\Phi)^2$ or $\frac{1}{6}(-\be\Phi)^3$ (see~\cite{b13}). We stress that the neutrality of the hydrogen atom plays an important role in the estimations (see cases (a) and (c)). In particular, when the root cluster $C(N\pr,N\el)$ is the electron-proton cluster $C(1,1)$, a rough estimation of its contribution based on~\eqref{estim g} would give $\be e^2 \rho\atH\ide/k^2$, which is not small compared to $\rho\atH\ide$ when $k$ is in the dielectric range (namely \eqref{neglige ionic} no longer applies). But in fact, as shown in details in section~\ref{P atomic contribution}, the root cluster $C(1,1)$ is really of the order $\rho\atH\ide\alphaH$ in the dielectric regime: the dangerous factor $1/k^2$ is cancelled by neutrality. More generally, the estimate~\eqref{estim g} is not optimal when $C(N\pr,N\el)$ is a neutral root cluster ($N\pr=N\el$) corresponding to a molecular state of $N\pr$ protons and $N\el$ electrons. Its contribution does not behave as $\be e^2 \rho_{N\pr N\el}\ide/k^2$, as estimated in~(a), but rather as $\rho_{N\pr N\el}\ide \al_{N\pr N\el}$ with $\al_{N\pr N\el}$ the polarisability of the molecule. If $C(N\pr,N\el)$ is not neutral, we cannot improve on the estimate $\be e^2 \rho\ide_{N\pr N\el}/k^2$ and hence the use of \eqref{Strong inequ} cannot be avoided when $\rho_{N\pr N\el}\ide$ is the density of a charged ion.

\paragraph{Graphs with two root clusters} We consider eventually the case of graphs containing two root clusters $C_a$ and $C_b$. Since we perform in \eqref{SCE rho(L_a,L_b)} an integration over $C_a$, the latter cluster can be treated as a black cluster with a special weight $Z^{\TT*}_\phi(C_a)$. Writing $e_{\al_a}e_{\al_b}=e^2 \text{sgn}(e_{\al_a}) \text{sgn}(e_{\al_b})$ in \eqref{A2 chiTk loop}, all graphs obviously have a prefactor $\be e^2 / k^2$, as before. The special weight $Z^{\TT*}_\phi(C_a)$ differs from $Z^{\TT}_\phi(C_a)$ by an additionnal factor $\exp[-i\vk\cdot\R_i]\text{sgn}(e_{\al_i})q_i$ for the loop $\LL_i$ identified to $\LL_a$. These factors do not depend on~$\be$, and change the estimations only in one case: the pseudo-neutrality condition cannot be used anymore, because of the factor $\text{sgn}(e_\al)$, to cancel the contributions of order~$z^0$ associated to clusters $C_a(1,0)$ or $C_a(0,1)$ connected by a single bond $-\be\Phi$ by adding these two contributions. However, because of the factor $\exp[-i\vk\cdot\R]$, the integration over the cluster $C_a$ involves the Fourier transform of the bond $-\be\phi$, which is of order
\eq{	\label{estimation fin}
     \be e^2 \rho\el\ide/(k^2+\kappa^2).
}
When $k$ is large enough for \eqref{neglige ionic} to hold, the graphs with two root clusters are hence all negligible in the atomic limit.

%As we have seen in the evaluation of the graphs~\eqref{4 o-o}, the integration over the cluster $C_a$ involves however the Fourier transform of the bond $-\be\phi$, because of the factor $\exp[-i\vk\cdot\R]$, and these contributions, which are of order
%[...]
%are hence nevertheless individually negligible since $k$ is large enough for \eqref{neglige ionic} to hold. %It is clear from the definitions of the weights $Z_{\CC^{(a)}}$ and $Z_{\CC^{(b)}}$ that the neutrality of the hydrogen atom does play in these graphs the same role as in the graphs with a single root cluster.

In conclusion, we have shown that if $k=k_\be$ remains in the range $\lam_{\rm I}^{-1}\ll k\ll \lam\atH^{-1}$, the only graph giving a non vanishing contribution to $\chiTk/\rho\atH\ide$ in the limit $\be \to \infty$ is the pure atomic graph $C(1,1)$ calculated in section~\ref{P atomic contribution}.

\section{On the cross-over regime} 	\label{S cross-over}

When $k$ is not large in front of $\lam_{\rm I}^{-1}$, ionic screening cannot be neglected, and $\chiTk$ interpolates between the values $\TF{\chi}(\vz)=-1$ and $-4\pi\rho\atH\ide \alphaH$. This cross-over regime between ionic and dielectric screening can be described approximately by adding further diagrams to the graphs calculated in the previous section. We noted earlier that the four graphs \eqref{4 o-o}, which give the mean-field result \eqref{MF result}, saturate the perfect screening relation \eqref{A2 -1}. Taken individually, the other graphs in the screened cluster expansion do not give however contributions to $\chiTk$ that vanish when $k\to 0$. Contributions compatible with perfect screening are obtained by ``dressing'' the root points with a cluster $C(1,0)$ or $C(0,1)$ connected by a bond $-\be\Phi$, similarly to the ``dressing'' procedure used to prove the screening sum rules obeyed by the particle correlations using their Mayer-loop expansions~\cite{BMA}.

\begin{figure}[ht]
\begin{center} \begin{small}
\psfrag{chi}{$\chiTk$}
\psfrag{-1}{$-1$}
\psfrag{10-7}{{\footnotesize $-10^{-7}$}}
\psfrag{2 10-7}{{\footnotesize $-2\cdot10^{-7}$}}
\psfrag{alpha}{$-4\pi\rho\atH\ide\alphaH$}
\psfrag{k}{$k$}
\psfrag{lami}{$\lam_{\rm I}^{-1}$}
\epsfig{file=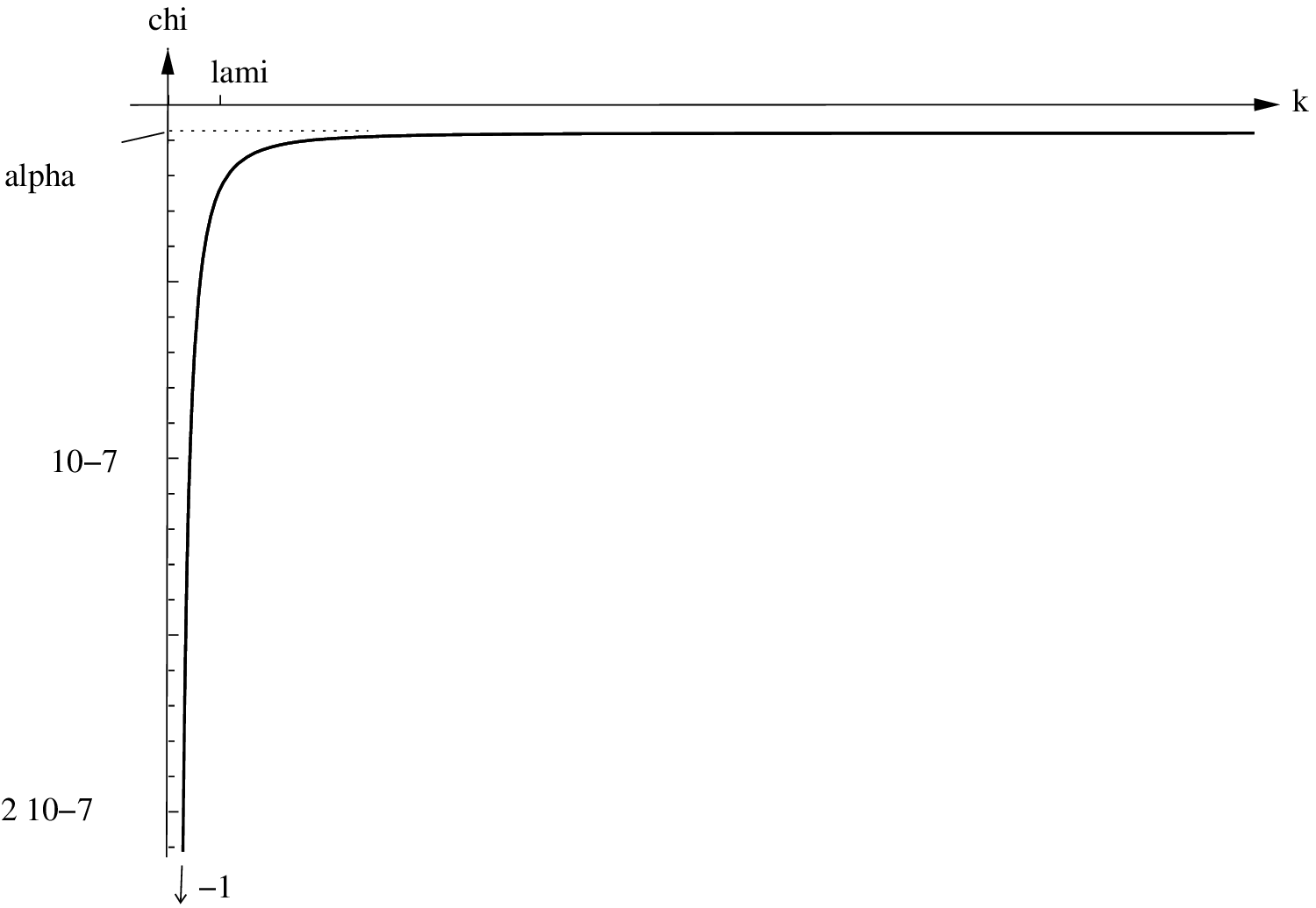}
\caption{\label{fig. plot}
{\footnotesize   Plot of the response function~$\chiTk$, according to~\eqref{approx chi}.   The electron-proton plasma is in an dilute atomic phase of density $\rho\atH\ide=10^{15}$ atoms/cm$^{3}$. The temperature is $T=1350$~K, and the Saha degree of ionization $\gamma\equiv\rho\el\ide/(\rho\el\ide+\rho\atH\ide)=10^{-24}$. In this situation, the cross-over distance \eqref{A2 lami} between ionic and dielectric screening takes the value $\lam_{\rm I}\simeq 700$~cm. On the plot, the inverse Debye screening length $\kappa \simeq 10^{-7}$~cm$^{-1}$ is almost at the origin, and the inverse lengths $1/(\be e^2)\ll\lam^{-1}\atH \simeq 10^{9}$~cm$^{-1}$ are far to the right. This plot of $\chiTk$ shows the plateau of this function for $k$ in the range $\lam_{\rm I}^{-1}\ll k \ll \lam\atH^{-1}$, which corresponds to the dielectric response of a gas of atomic hydrogen.
}
   }
    \end{small}  \end{center}
\end{figure}

We can obtain a model that describes the cross-over regime between ionic and dielectric screening, by adding to $\chiTMFk$ the contributions of the ``dressed'' hydrogen cluster $C(1,1)$:
\eq{	\label{C11 hab}
      \epsfig{file=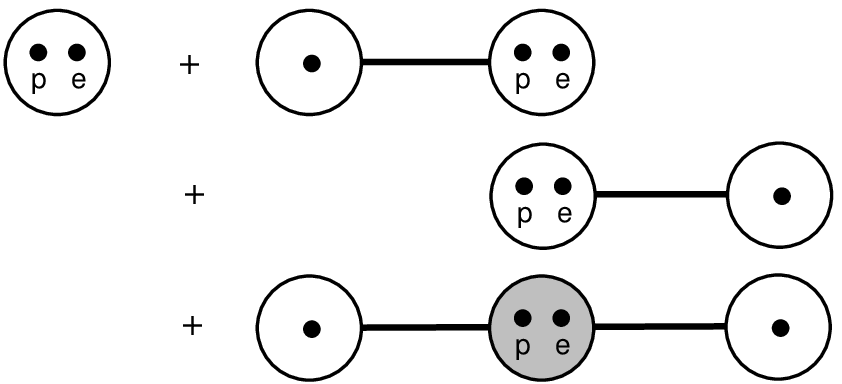}
}
where the unspecified particle in the root cluster $C_a$ and $C_b$ can be either a proton or an electron. Let us evaluate at leading order the second graph in~\eqref{C11 hab}, for $k \ll \lam^{-1}$. Its contribution can be represented by the eight diagrams\footnote{The representation \eqref{Hab gauche o-} is in fact nothing but the set of the loop-Mayer graphs in the expansion of $\rho_{\TT}(\LL_a,\LL_b)$ that have been collected together in \cite{b13} to form the considered screened cluster graph.}
\eq{	\label{Hab gauche o-}
   \Graph{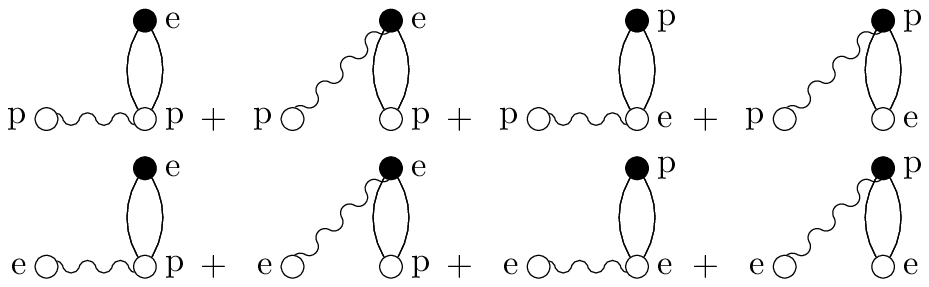}
}
where the vertices represent loops $\LL=(\R,\al,q=1,\vxi)$ of weight $z(\LL)=2 z_\al /(2\pi\lam_\al^2)^{3/2}$. The species of each loop ($\al=\text{e}$ or~p) is indicated on the graphs. The wiggly line represents the bond $-\be\phi$, and the root loop $\LL_b$ and the internal loop are connected by the bond ${\cal B}_{\phi,2}^{\TT}$. In every graph of~\eqref{Hab gauche o-}, the integration over $\LL_a$ (see \eqref{A2 chiTk loop}) involves the evaluation of the Fourier transform of the bond $-\be\phi$:
\eq{	\label{int o-}
   \sum_{\al_a=\text{e,p}} e_{\al_a} \frac{2 z_{\al_a}}{(2\pi\lam^2_{\al_a})^{3/2}} \int\romD(\vxi_a) \int\romd\R_a \, \e{-i\vk\cdot\R_a}  (-\be\phi(\LL_a,\LL_i))
}
where $\LL_i$ is the white loop $\LL_b$ or the black loop. At leading order, \eqref{int o-} can be calculated by retaining only the contribution of the term $n=0$ in~\eqref{A2 TF phi}\footnote{The part $\phi^{[n\neq 0]}(\LL_a,\LL_b)$ associated to the non zero frequency terms in~\eqref{A2 TF phi} is integrable at zero density (see~\cite{BMA}), so that its contribution to~\eqref{int o-} is $\OO(z)$ for all values of~$k$ (discarding powers of~$\be$).}, with the result:
\eq{
    -\frac{\kappa^2(k)}{k^2 + \kappa^2(k)} e_{\al_i} \e{-i\vk\cdot\R_i} \int_0^1\romd t\, \e{-i\vk\cdot\lam_i\vxi_i(t)}.
}
The contribution of the eight graphs~\eqref{Hab gauche o-} to $\chiTk$ is hence
\begin{multline}	\label{A2 atomic ecrante}
    -\frac{4\pi\be e^2}{k^2} \frac{2z\el}{(2\pi\lam\el^2)^{3/2}}\frac{2z\pr}{(2\pi\lam\pr^2)^{3/2}}
    \frac{\kappa^2(k)}{k^2 + \kappa^2(k)}
    \int\romd\ver \int\romD(\vxi\el)\romD(\vxi\pr)	\\
    \int_0^1\romd s\int_0^1\romd t \,
    	 {\cal B}_{\phi,2}^{\TT}(\ver,\chi\el,\chi\pr)
    	 \Big(
    	\e{i\vk\cdot\lam\el\vxi\el(s)}\e{-i\vk\cdot\lam\el\vxi\el(t)} +
	\e{i\vk\cdot\lam\pr\vxi\pr(s)} \e{-i\vk\cdot\lam\pr\vxi\pr(t)}	\\
    	-\e{-i\vk\cdot\ver} \e{i\vk\cdot\lam\pr\vxi\pr(s)} \e{-i\vk\cdot\lam\el\vxi\el(t)}
    	-\e{i\vk\cdot\ver}  \e{i\vk\cdot\lam\el\vxi\el(s)} \e{-i\vk\cdot\lam\pr\vxi\pr(t)}
    	\Big)
    	 .
\end{multline}
This expression is similar to \eqref{A2 98} and can be calculated at low temperature in the same way (see appendix~B). According to~\eqref{B.resultat}, its leading term when $k\ll \lam^{-1}$ and $\be \to \infty$ is also related to the polarisability~$\alphaH$ of the ground state of the hydrogen atom:
\eq{
    -4\pi\rho\atH\ide\alphaH \Big( \frac{-\kappa^2}{k^2 + \kappa^2} \Big),
}
where we used $\kappa^2(k)\simeq \kappa^2$. Using~\eqref{super prop}, the third graph in~\eqref{C11 hab} is easily seen to give a contribution identical to that of the cluster $C(1,1)$, with an additional factor $-\kappa^2(k)/(k^2+\kappa^2(k))$ due to the ``dressing'' by the root cluster $C_b=C(1,0)$ or $C(0,1)$. The last graph in~\eqref{C11 hab} involves similarly two such dressing factors. We find therefore, at leading order and for $k\lam\ll 1$,
\eq{	\label{APPROX}
    \TF{\chi}_{11}(\vk) \simeq -4\pi\rho\atH\ide \alphaH \Big[ 1- \frac{2\kappa^2}{k^2+\kappa^2} + \Big( \frac{-\kappa^2}{k^2+\kappa^2}\Big)^2 \Big] \simeq -4\pi\rho\atH\ide \alphaH \Big( \frac{k^2}{k^2+\kappa^2}\Big)^2.
}
This expression vanishes at $k=0$ thanks to the screening factors $k^2/(k^2+\kappa^2)$. Adding \eqref{APPROX} to \eqref{MF result}, we obtain the following approximation to $\chiTk$ in the atomic limit
\eq{	\label{approx chi}
   \chiTk \simeq -\frac{\kappa^2}{k^2 + \kappa^2} - 4\pi\rho\atH\ide \alphaH \Big( \frac{k^2}{k^2+\kappa^2}\Big)^2,
   \qquad k\lam_\al\ll 1.
}
Eq.~\eqref{approx chi} describes simultaneously the screening due to free electrons and protons (first term), and the screening due to the polarisability of the hydrogen atoms. A plot of~\eqref{approx chi} is shown on Fig.~\ref{fig. plot}.
%________________________________________ figure
%_________________________________________

The approximation~\eqref{approx chi} can be compared with the extended RPA-dielectric function introduced by R\"{o}pke and Der~\cite{RD}. We first note that, in the Maxwell-Boltzmann approximation, the RPA-dielectric function
\eq{
   \eps_{\rm RPA}(\vk) \simeq -\frac{4\pi}{k^2} \sum_{\al=\text{e,p}} \frac{2 e_{\al}^2}{(2\pi)^3}
   \int\romd\vp\, \frac{\e{-\be(E_{\vp}^{[\al]}-\mu_\al)} - \e{-\be(E_{\vp-\vk}^{[\al]}-\mu_\al)}}{E_{\vp}^{[\al]} - E_{\vp-\vk}^{[\al]}},
}
$E_{\vp}^{[\al]}=\hbar^2 \vp^2/(2 m_\al)$, is simply related to the function $\kappa^2(k)$. If we replace the factor \eqref{A2 f_tau(k)} in \eqref{A2 def kappa2(k)} by \eqref{A2 int dp} and perform the ``time'' integral, we find indeed $\eps_{\rm RPA}(\vk)=\kappa^2(k)/k^2$. Using~\eqref{A2 2}, $\eps_{\rm RPA}(\vk)$ is therefore equivalent to the mean-field response function~\eqref{MF result}:
\eq{
   \frac{1}{\eps_{\rm RPA}(\vk)} - 1 = \chiTMFk.
}
The extended RPA-dielectric function includes a contribution $\eps_{\rm at}(\vk)$ associated to the polarisability of the hydrogen atom:
\eq{	\label{approx RD}
   \eps(\vk) = 1 + \eps_{\rm RPA}(\vk) + \eps_{\rm at}(\vk)
}
with
\eq{	\label{def eps_at}
   \eps_{\rm at}(\vk) \simeq  -\frac{4\pi e^2}{k^2}  \frac{4\e{\be\mu}}{(2\pi)^3}
   \int\romd\vp \sum_{\substack{\n,\n'=0\\\n\cdot\n'=0}}^{\infty} |A_{\n\n'}(\vk)|^2\, \frac{\e{-\be(E_{\n}+E_{\vp})} -  \e{-\be(E_{\n'}+E_{\vp-\vk})}}{E_{\n} + E_{\vp} - E_{\n'} - E_{\vp-\vk}}
}
($E_{\n}$, $E_{\vp}$ and $|A_{\n\n'}(\vk)|^2$ are defined in section~\ref{S elementary}). The above expression of $\eps_{\rm at}(\vk)$ follows from eq. (2.4) of~\cite{RD} (see also eq. (4.265) in~\cite{KKER}) by setting $\omega=0$, neglecting degeneracy effects, and retaining only the terms involving at least one ground state (because $\e{-\be E_0} \gg \e{-\be E_{\n}}$, $\n \neq 0$). Comparing \eqref{def eps_at} with \eqref{def chi_at^0} and \eqref{A2 int dp}, we see that $\eps_{\rm at}(\vk)$ is the same as our expression $\TF{\chi}_{\rm at}^{(0)}(\vk)$ \eqref{A2 106}. The approximations \eqref{approx chi} and \eqref{approx RD} for the response function are therefore equivalent at leading order: they differ only by a term $\propto (\rho\atH\ide \alphaH)^2$. R\"{o}pke and Der derived $\eps_{\rm at}(\vk)$ in the framework of the Feynman perturbation theory of the many-body problem. They must hence sum an infinite number of graphs to obtain the atomic contribution, since chemical binding is a non-linear effect in the interaction. This is in marked contrast with our non-perturbative approach, where the atomic contribution (and the contributions of other bound states) are described by simple graphs.

We stress that the approximation \eqref{approx chi} does not include in a systematic way the leading corrections to $\chiTMFk$ for all values of $k\ll \lam_{\rm I}^{-1}$, even in an atomic phase satisfying the inequalities \eqref{Strong inequ}. Consider for example the contribution of the graph made of a root cluster $C(1,2)$ describing a H$^-$ ion. According to \eqref{def g} and \eqref{estim g}, this graph will lead, when suitably dressed, to a term of order
\eq{	\label{estim C_12}
   \parenth{\frac{k^2}{k^2 + \kappa^2}}^2 \frac{\be e^2 \rho_{12}\ide}{k^2}.
}
When $k\simeq \kappa$, \eqref{estim C_12} is hence of the order of a polynomial in~$\be$ times $\rho_{12}\ide/\rho\el\ide$ (since $\kappa^2\propto z$). Although $\rho_{12}\ide/\rho\el\ide\ll 1$ in an atomic phase satisfying~\eqref{Strong inequ}, it is much larger than $\rho\atH\ide \alphaH$ when $\be\to \infty$, as can be seen by inserting the values of the binding energies $E\atH\simeq -13.6$~eV and $E_{12}\simeq -14.3$~eV. The second term in the approximation \eqref{approx chi} is therefore not the leading correction to $\chiTk-\chiTMFk$ when $k\simeq \kappa$. Obtaining a coherent approximation for $\chiTk$ at low densities for all values of~$\vk$ is thus a subtle problem, since all ions contribute to the constitution of the screening cloud when $k$ is in the cross-over region.\\

In conclusion, the screened cluster expansion is a convenient technique to study at low densities the response function (and other static equilibrium quantities) of the quantum plasma when the charges recombine into bound entities. With this technique, we have been able to determine the wave length range where $\chiTk$ shows a plateau corresponding to dielectric behaviour. For $k$ in this range, we have calculated the dielectric constant up to first order in the atomic density, taking into account in a controlled way all effects induced by the Coulomb potential (chemical binding, collective screening, polarization). The present method allows in principle quantitative calculations of other type of contributions, coming from higher order density terms, thermal excitations, more complex chemical species, etc. Finally, it can be extended to the study of more general partially ionized nucleo-electronic plasma.

\appendix

\section{Upper bounds on remainders}

In this appendix, we prove that the various terms we have discarded in section~6 are indeed negligible in the atomic limit. The points (i) and (ii) are established in section A.1 and A.2 respectively. Section A.3 contains a few lemmas which are used in the proofs. The letter~$C$ denotes throughout this appendix a positive constant which can have different values at different places.

\subsection{Neglecting screening effects in the atomic contribution}

We seek for upper bounds on the contributions to~$\chiTk$ of the terms~$R_i$ (see~\eqref{A2 def R_1}-\eqref{A2 def R_3}), which arise when the screened potential is replaced by the bare Coulomb potential in~\eqref{A2 98}. These contributions have the generic form
\begin{multline}	\label{A2 zxcv}
    -\frac{4\pi\be e^2}{k^2} \int\romd\ver  \int\romD(\vxi\el)\int\romD(\vxi\pr)\int_0^1\romd s
    	\frac{2 z\el}{(2\pi\lam\el^2)^{3/2}}\frac{2 z\pr}{(2\pi\lam\pr^2)^{3/2}}
    	 \, \e{I_R(\LL\el)} \e{I_R(\LL\pr)} \\ \times
    	\e{i\vk\cdot\lam\pr\vxi\pr(s)} (\e{i\vk\cdot[\ver+\lam\el\vxi\el(s)-\lam\pr\vxi\pr(s)]}-1)
    	(\e{-i\vk\cdot\ver}-1)   \,R(\ver,\chi\el,\chi\pr),
\end{multline}
where $R = R_1, R_2, R_3$. We introduce absolute values, use $|\e{ix}-1| \leq C |x|$ and $|\exp[-\be I_R(\LL)]|\leq C$ (this follows from \eqref{def z_phi(LL)} and~\eqref{A2 borne phich}). Expression~\eqref{A2 zxcv} can be majorized by
\begin{multline}	\label{A2 contrib R}
    C \be e^2
     	\frac{2 z\el}{(2\pi\lam\el^2)^{3/2}}\frac{2 z\pr}{(2\pi\lam\pr^2)^{3/2}}
     	 \int\romd\ver  \int\romD(\vxi\el)\int\romD(\vxi\pr)\int_0^1\romd s  |R(\ver,\chi\el,\chi\pr)| \\ \times
     	 |\ver + \lam\el\vxi\el(s)-\lam\pr\vxi\pr(s)| |\ver|.
\end{multline}
We study first the contributions of the terms $R_2$ and $R_3$, and then that of~$R_1$.

\subsubsection*{Contribution of $R_2$}
%%%%%%%%%%%%%%%%

Inserting~\eqref{A2 def R_2} in~\eqref{A2 contrib R}, we have to estimate a finite sum of terms of the form
\begin{multline}	\label{A2 contrib R_2}
    P(\be) z\el z\pr 
     	 \int\romd\ver  \int\romD(\vxi\el)\int\romD(\vxi\pr)\int_0^1\romd s
     	 |\phich(\ver,\chi\el,\chi\pr)|^n |V(\ver,\chi\el,\chi\pr)|^m \\ \times
     	 |\ver + \lam\el\vxi\el(s)-\lam\pr\vxi\pr(s)| |\ver|
\end{multline}
where $P(\be)$ is a polynomial in~$\be$ and $n+m\geq 6$. It is proven in~\cite{BMA} that, at low density, the potential~$\phich$ can be majorized by a constant times the electrostatic potential~\eqref{A2 def V_elec(LL_i,LL_j)}
\eq{
    |\phich(\ver,\chia,\chib)| \leq C |V_{\text{elec}}(\ver,\chia,\chib)|.
}
Using this majoration, and introducing the change of variables $\ver=\lam\vx$, we obtain the following upper bound for~\eqref{A2 contrib R_2}:
\begin{multline} \label{A2 borne sup R_2}
        P(\sqrt{\be}) z\el z\pr 
          \int\romd\vx  \, |\vx| \int\romD(\vxi\el)\int\romD(\vxi\pr)
        \Big(   \int_0^1\romd s\, \Big|\vx +\sqrt{\frac{m}{m\el}}\vxi\el(s)-\sqrt{\frac{m}{m\pr}}\vxi\pr(s) \Big| \Big) 
	\\	\times
     	 \Big(\int_0^1\romd\tau_a\int_0^1\romd\taub
     	   \frac{1}{|\vx + \sqrt{\frac{m}{m\el}}\vxi\el(\taua)-\sqrt{\frac{m}{m\pr}}\vxi\pr(\taub)|} \Big)^n \\ \times
          \Big(\int_0^1\romd\tau
     	 \frac{1}{|\vx + \sqrt{\frac{m}{m\el}}\vxi\el(\tau)-\sqrt{\frac{m}{m\pr}}\vxi\pr(\tau)|} \Big)^{m}
\end{multline}
Let us show that the above integrals are finite, thereby proving that \emph{the contribution of~$R_2$ is dominated by a polynomial in~$\sqrt{\be}$ times $z\el z\pr$}.

Using twice the Cauchy-Schwartz inequality on the measure $\romD(\vxi\el)\romD(\vxi\pr)$, one can majorize the integrals in~\eqref{A2 borne sup R_2} by $\int\romd\vx J(\vx)$, where
\begin{multline}	\label{A2 C4}
    J(\vx) \equiv |\vx|
    \sqrt{\int\romD(\vxi) \Big(  \int_0^1\romd s\, |\vx + \vxi(s)| \Big)^2 } \times
     \bigg[  \int\romD(\vxi)   \Big(\int_0^1\romd s\,
        \frac{1}{|\vx+\vxi(s)|} \Big)^{4m}   \\ \times    
      \int\romD(\vxi\el)\int\romD(\vxi\pr) \Big(\int_0^1\romd\tau_a\int_0^1\romd\taub
        \frac{1}{\left|\vx+\sqrt{\frac{m}{m_a}}\vxi\el(\taua) - \sqrt{\frac{m}{m_b}} \vxi\pr(\taub)\right|}\Big)^{4n} 
        \bigg]^{\frac{1}{4}}.
\end{multline}
In~\eqref{A2 C4}, we have used~\eqref{A2 Coord cm/rel} to write the functional integrals involving a single time in terms of the relative path $\vxi(s) = \sqrt{{m}/{m\el}}\vxi\el(s)-\sqrt{{m}/{m\pr}}\vxi\pr(s)$. The following lemma can be used to prove the finiteness of the integral over~$J(\vx)$.
%%%%%%%%%%%%%%%%%%%%%%%%%%%%%%%%%%%%%%%% LEMMA 1
\begin{lemma}	\label{A2 lemme smooth}
(smoothing the Coulomb singularity)
\begin{gather}
\label{A2 149}
    a)\quad \int\romD(\vxi)\, \Big( \int_0^1\romd s\, \frac{1}{|\vx + \vxi(s)|} \Big)^m 
    	\leq \frac{C}{|\vx|^m+1}	\\
\label{A2 smooth}
    b)\quad     \int\romD(\vxi_a)\int\romD(\vxi_b) \Big(\int_0^1\romd\tau_a\int_0^1\romd\taub
    	\frac{1}{|\vx + \sqrt{\frac{m}{m_a}}\vxi_a(\taua)-\sqrt{\frac{m}{m_b}}\vxi_b(\taub)|}\Big)^n \leq \frac{C}{|\vx|^n + 1}.
\end{gather}
 \end{lemma}
%%%%%%%%%%%%%%%%%%%%%%%%%%%%%%%%%%%%%%%%%%
The functional integrations over the Brownian paths smooth out the Coulomb singularity at the origin so that \eqref{A2 149} and~\eqref{A2 smooth} remain finite when $|\vx|\tend 0$. Note furthermore that
\eq{
    \int\romD(\vxi)\, \Big( \int_0^1\romd s\, |\vx + \vxi(s)| \Big)^2 \leq C|\vx|^2.
}
The function~$J(\vx)$ can therefore be majorized by $J(\vx) \leq C |\vx|^2/(|\vx|^{4(n+m)}+1)^{1/4}$, and $\int\romd\vx J(\vx)$  is hence integrable since $n+m\geq 6$. The contribution of~$R_2$ to~$\chiTk$ is thus indeed dominated by a polynomial in~$\sqrt{\be}$ times $z\el z\pr$.

\subsubsection*{Contribution of $R_3$}
%%%%%%%%%%%%%%%%

We seek for an upper bound for
\begin{multline}
    P(\be) z\el z\pr 
     	 \int\romd\ver  \int\romD(\vxi\el)\int\romD(\vxi\pr)\int_0^1\romd s\,
     	 |\ver + \lam\el\vxi\el(s)-\lam\pr\vxi\pr(s)| |\ver| \\ \times
     	 \sum_{n=6}^\infty \frac{1}{n!}(\be |\phich(\ver,\chi\el,\chi\pr)|)^n 
     	 \sum_{m=0}^5 (\be |V(\ver,\chi\el,\chi\pr)|)^m.
\end{multline}
We write $|\phich|^n = |\phich|^{n-6} |\phich|^6$ and, using~\eqref{A2 borne phich}, majorize the sum over~$n$ by
\eq{
    \sum_{n=6}^\infty \frac{1}{n!} \parenth{\be |\phich(\ver,\chi\el,\chi\pr)| }^{n-6}
    \leq \e{\be  |\phich|} \leq C.
}
After this majoration, we are left with an integral already studied (see~\eqref{A2 contrib R_2}), so that the contribution of~$R_3$, like that of~$R_2$, is dominated by a polynomial in~$\sqrt{\be}$ times $z\el z\pr$.

\subsubsection*{Contribution of $R_1$}
%%%%%%%%%%%%%%%%

Since $|R_1(\ver,\chi\el,\chi\pr)|$ can be majorized by $C\, \be e^2 \kappa\, \BB_{V,2}^{\TT\TT}(\ver,\chi\el,\chi\pr)$, using \eqref{A2 def R_1}, \eqref{A2 borne phich} and the positivity of $\BB_{V,2}^{\TT\TT}(\ver,\chi\el,\chi\pr)$ (see~\eqref{A2 pos W_V^TT}), we expect its contribution to~$\chiTk$ to be smaller by a factor $\be e^2 \kappa$ as compared to the leading contribution $\TF{\chi}_{\rm at}(\vk)$~\eqref{A2 111}. Writing~\eqref{A2 contrib R} in the center of mass and relative coordinates, we find that the contribution of~$R_1$ is dominated by $C\,\be e^2\kappa\, (2\pi\lamat^2)^{-3/2}\exp[2\be\mu]$ times
\eq{	\label{A2 130}
    \frac{\be e^2}{(2\pi\lam^2)^{3/2}} \int\romd\ver\int\romD(\vxi)\int_0^1\romd s\,
    |\ver+\lam\vxi(s)| |\ver| \, \BB_{V,2}^{\TT\TT}(\ver,\lam\vxi).
}
We show here that~\eqref{A2 130} behaves like a polynomial in~$\sqrt{\be}$ times $\exp[-\be E_0]$, just as~\eqref{A2 111}. \emph{The contribution of~$R_1$ is hence dominated by} $P(\sqrt{\be}) \, (\be e^2\kappa)\, \rhoat$, which is exponentially smaller than~$\chiTatk$.

In order to prove that~\eqref{A2 130} behaves like $\exp[-\be E_0]$ at low temperatures, one must convert the functional integral into operator's language to extract the ground state energy~$E_0$. The main difficulty in the proof is then to keep track of the convergence of the $\ver$-integral, which is obvious in the functional integral representation, but is non trivial in the operatorial expressions. In the language of operators, \eqref{A2 130} is equivalent to
\eq{	\label{A2 superbe}
    e^2 \int_0^\be \romd\tau\, \Tr\Big\{
    \opU(\be-\tau)|\opq|\opU(\tau)|\opq| - \e{-\be\opH_0} \opT \Big[
    |\opq|(\tau) |\opq| \sum_{n=0}^5\frac{1}{n!} \Big( \int_0^\be\romd s \overline{\opV}(s) \Big)^n \Big] \Big\},
}
where we use the same notation as in~\eqref{A2 OKOK}. From the Dyson series, we have
\eq{	\label{A2 Dyson}
    \opU(t_2-t_1) = \e{-(t_2-t_1)(\opH+\opV)} = \exp[-t_2 \opH_0] \opT  \exp \Big[ -\int_{t_1}^{t_2}\romd s\, \opV(s) \Big] \exp[t_1 \opH_0],
}
so that~\eqref{A2 superbe} can be rewritten as
\eq{
    e^2 \int_0^\be \romd\tau\, \Tr\Big\{
    \e{-\be\opH_0} \opT \Big[ |\opq|(\tau) |\opq|
    \Big( \e{ -\int_0^\be \romd s\, \opV(s)} -
     \sum_{n=0}^5\frac{1}{n!} \Big( \int_0^\be\romd s \overline{\opV}(s) \Big)^n
     \Big) \Big] \Big\}.
}
The truncation terms in~$\BB_{V,2}^{\TT\TT}$ therefore subtract out of the trace the terms of order 0 to~5 in~$\opV$, ensuring thereby the finiteness of the trace (the same argument shows that the trace~\eqref{A2 OKOK} converges). To exploit these cancellations, we introduce in~\eqref{A2 superbe} a \emph{limited} Dyson expansion of~$\opU(s)$ to fifth order in~$\opV$:
\eq{	\label{A2 Dyson5 U(s)}
    \opU(s) = \sum_{k=0}^5 \opU^{(k)}(s) + \opU\Reste(s),
}
where $\opU^{(0)}(s) = \opU_0(s) =  \e{-s \opH_0}$ is the free propagator,
\eq{	\label{A2 def U^(k)}
    \opU^{(k)}(s) = \opU_0(s) \frac{1}{k!} \opT \Big(  \int_0^s\romd s'\, \overline{\opV}(s') \Big)^k,
    	\qquad k=0,1,\ldots,5,
% ancienne forme classique pour ce developpement:    
%     (-1)^k\! \int_0^s\!\romd s_1\int_0^{s_1}\!\romd s_2\ldots
%    \int_0^{s_{k-1}}\!\!\romd s_k\, \opU^{(0)}(s) \opV(s_1) \opV(s_2) \ldots \opV(s_k)
}
with $\overline{\opV}=-\opV=e^2/|\opq|$, and the remainder is
\eq{	\label{A2 def U^R}
    \opU\Reste(s) =  \int_0^s\!\romd s_1\int_0^{s_1}\!\romd s_2\ldots
    \int_0^{s_5}\!\romd s_6\, \opU_0(s) \overline{\opV}(s_1) \ldots \overline{\opV}(s_5)
    \opU_0(s_6) \overline{\opV} \opU(s_6).
}
Notice that the $\opU^{(k)}(s)$ involve only free propagators $\opU_0(s)$ (and Coulomb operators~$\opV$), while $\opU\Reste(s)$ contains the full evolution operator~$\opU(s)$.
Because of the truncation, the terms in~\eqref{A2 superbe} of order less than~6 in~$V$ cancel out, and we are left with terms of the form
\eq{	\label{A2 xx}
    e^2 \int_0^\be \romd \tau\, \Tr \Big\{ \opU^{(l)}(\be-\tau) |\opq| \opU^{(m)}(\tau) |\opq| \Big\},
}
where $l,m=\text{D}$ or $0,1,2,3,4,5$ with $6\leq l+m\leq 10$ if $l,m\neq\text{D}$.

Consider first the case where $l$ and $m$ are different from~$\text{D}$. Since only free propagators $\opU_0$ are present, no bound state can occur in these terms. Let us show that~\eqref{A2 xx} is then bounded by a polynomial in~$\sqrt{\be}$, implying a bound $P(\sqrt{\be}) z\el z\pr$ for the contribution of these terms to~$\chiTk$.
Inserting~\eqref{A2 def U^(k)} into \eqref{A2 xx}, we find
\begin{multline}	\label{A2 c1}
    \frac{e^2}{l!m!} \int_0^\be\romd\tau \int_\tau^\be\romd s_1\ldots\romd s_l
    \int_0^\tau\romd s_1' \ldots \romd s_m' \\
    \Tr \Big\{
    \opU_0(\be)\opT \Big[\overline{\opV}(s_1) \ldots \overline{\opV}(s_l) |\opq|(\tau)\overline{\opV}(s_1')\ldots\overline{\opV}(s_m')|\opq|
    \Big] \Big\},
\end{multline}
\newcommand{\opVB}{\overline{\opV}}
which can be majorized by
\eq{	\label{A2 c2}
    \frac{e^2}{l!m!} \int_0^\be\romd\tau \int_0^\be\romd s_1\ldots \romd s_{l+m}\,
    \Tr \Big\{
    \opU_0(\be)\opT \Big[ |\opq|(\tau) |\opq| \opVB(s_1) \ldots \opVB(s_{l+m})
    \Big] \Big\},
}
by extending all the integrals to the domain~$[0,\be]$ (the integrand is positive, as can be seen from the functional integral representation of the trace). Expression \eqref{A2 c2} is equivalent to
\eq{	\label{A2 c3}
    \frac{(\be e^2)^{l+m+1}}{l!m!}\frac{1}{(2\pi\lam^2)^{3/2}} \int\romd\ver
    \int\romD(\vxi)\int_0^1\romd\tau\, |\ver+\lam\vxi(\tau)|\, |\ver|\,
    \Big( \int_0^1\romd s\, \frac{1}{|\ver+\lam\vxi(s)|} \Big)^{l+m}.
}
By the same argument (based on lemma \ref{A2 lemme smooth}) already used in estimating the contribution of~$R_2$, we conclude that \eqref{A2 c3} is bounded by a polynomial in~$\sqrt{\be}$.

Consider now the terms with at least one index equal to~D. We use the invariance of the trace under cyclic permutations and the fact that $|\Tr\opA| \leq \Vert \opA \Vert_1$ (where $\Vert \cdot \Vert_1$ is the trace norm), to majorize~\eqref{A2 xx} by
\eq{	\label{A2 141}
    e^2 \int_0^\be\romd\tau \, \Vert \, |\opq| \opU^{(l)}(\be - \tau) |\opq| \opU^{(m)}(\tau) \Vert_1.
}
One can obtain the following results on the norms of the operators
%%%%%%%%%%%%%%%%%%%%%%%%%%%%%%%%%%%%%%%%%%%% LEMME 2
\begin{lemma}	\label{A2 lemme 2}
(Convergence of the traces)
\begin{gather}
\label{A2 2a}
    a)\quad \Vert \, |\opq| \opU^{(l)}(s) \Vert \leq P(\sqrt{s}), \qquad l=1,2,\ldots,5	\\
\label{A2 2b}
    b)\quad \Vert \, |\opq| \opU^{({\rm D})}(s) \Vert_1 \leq P(\sqrt{s}) \e{-s E_0}	\\
\label{A2 2c}
    c)\quad \Vert \, |\opq| \opU_0(t) |\opq| \opU^{({\rm D})}(s) \Vert_1 \leq P(\sqrt{s}) P(\sqrt{t}) \e{-s E_0}.
\end{gather}
\end{lemma}
%%%%%%%%%%%%%%%%%%%%%%%%%%%%%%%%%%%%%%%%%%%%

The operator norm in lemma~\ref{A2 lemme 2}.a holds only when $l\geq 1$, because $\opU^{(l=0)}(s)$ does not contain a Coulomb operator to counterbalance the presence of the unbounded operator~$|\opq|$. Naively speaking, the trace norms in \ref{A2 lemme 2}.b and \ref{A2 lemme 2}.c are finite because $\opU\Reste(s)$ is of order~6 in~$\opV$, which is enough to ensure convergence in 3-dimensional space even with two operators~$|\opq|$. Furthermore, the bounds \ref{A2 lemme 2}.b and \ref{A2 lemme 2}.c grow like $\exp[-s E_0]$ because $\opU\Reste(s)$ contains the full evolution operator $\opU(s)$ which involves bounds states.

From the upper bounds of lemma~\ref{A2 lemme 2} and the inequality $\Vert \op{S} \op{T} \Vert \leq \Vert \op{S} \Vert \cdot \Vert \op{T} \Vert_1$, one easily deduces that~\eqref{A2 141} is dominated by a polynomial in~$\sqrt{\be}$ times $\exp[-\be E_0]$, as announced after~\eqref{A2 130}. Indeed, if one index is equal to~D (say $m=\text{D}$), one uses~\eqref{A2 2c} if $l=0$, and \eqref{A2 2a} together with \eqref{A2 2b} if $l=1,2,\ldots,5$. We use also of course the fact that $\be-\tau$ and $\tau$ do not exceed~$\be$ and majorize $P(\sqrt{\be-\tau})$ and $P(\sqrt{\tau})$ by polynomials $P(\sqrt{\be})$. The same argument applies to the case $l=\text{D}$, $m\neq\text{D}$ by using the invariance under cyclic permutations in the trace~\eqref{A2 xx} to bring $\opU\Reste(\be-\tau)$ to the right of the expression. Eventually, the case $l=m=\text{D}$ is treated similarly by applying twice~\eqref{A2 2b} together with $\Vert \op{T} \Vert \leq \Vert \op{T} \Vert_1$.

\subsection{Neglecting excited and ionized states contributions}

From its definition~\eqref{A2 split chi_at = 0+1}, the contribution of excited and ionized states to~$\chiTk$ is
\begin{multline}	\label{A2 Mil}
    \TF{\chi}_{\rm at}^{(1)}(\vk) = -4\pi e^2 \e{2\be\mu} \int_0^\be\romd\tau\, f_{\tau/\be}(\vk)
        \Tr \Big\{ \opU(\be-\tau) \opQ \op{B}_{-\vk} \opU(\tau)  \opQ \op{B}_{\vk}  \\
        - \e{-\be \opH_0} \opT \Big[ \op{B}_{-\vk}(\tau)\op{B}_{\vk}
        \sum_{n=0}^5 \frac{1}{n!} \Big( \int_0^\be \romd s \overline{\opV}(s) \Big)^n \Big] \Big\}
\end{multline}
with
\eq{
    \op{B}_{\vk} \equiv({\e{i\vk\cdot\frac{m\el}{M}\opq} - \e{-i\vk\cdot\frac{m\pr}{M}\opq}})\frac{1}{|\vk|} = 
    \e{-i\vk\cdot\frac{m\pr}{M}\opq} \, (\e{i\vk\cdot\opq}-1) \, \frac{1}{|\vk|}.
}
Expression~\eqref{A2 Mil} involves a truncated trace similar to~\eqref{A2 superbe}, except for the additional operators~$\opQ$ that project the wavefunctions on the set of excited and ionized states. Notice that the operator~$\op{B}_{\vk}$ plays here a role similar to~$|\opq|$, because we look for an upper bound that is uniform in~$\vk$ and $\lim_{k \tend 0} \op{B}_{\vk}=i\hat{\vk}\cdot\opq$. The methods used to derive upper bounds on~\eqref{A2 superbe} can also be applied here with minor modifications (see below). The result is that $\TF{\chi}_{\rm at}^{(1)}(\vk)$ is dominated by $P(\sqrt{\be}) \rhoat \e{-\be(E_1-E_0)}$, where $E_1$ is the energy of the first excited state of the hydrogen atom. $\TF{\chi}_{\rm at}^{(1)}(\vk)$ is hence exponentially smaller than $\TF{\chi}_{\rm at}^{(0)}(\vk)$.

Following the same method as in section A.1, we introduce in~\eqref{A2 Mil} the limited Dyson expansion~\eqref{A2 Dyson5 U(s)} of $\opU(s)$. We distinguish two types of terms:
\begin{itemize}
\item[$\bullet$] Class I: terms with at least one $\opU\Reste(s)$.
\item[$\bullet$] Class II: terms without~$\opU\Reste(s)$ (these terms are of order 0 to 10 in~$\opV$).
\end{itemize}
Let us show that the terms in the class~I are dominated by $P(\sqrt{\be})(2\pi\lamat^2)^{-3/2}$ $\exp[-\be(E_1-2\mu)]$ and hence are $o(\rhoat)$. Majorizing the traces by trace norms (and using the cyclicity of the trace), these terms are dominated by
\eq{	\label{A2 **}
    C \frac{\e{2\be\mu}}{(2\pi\lamat^2)^{3/2}} \cdot \int_0^\be\romd\tau \,\Vert \op{B}_{\vk} \opU^{(l)}(\be-\tau) \opQ \opB_{-\vk} \opU^{(m)}(\tau) \opQ \Vert_1,
}
where at least $l$ or $m$ is equal to~D (the other index running from 0 to 5).%%%%%%%%%%%%%%%%%%%%%%%%%%%%%%%%%%%%%%%%%%%% LEMME 2
\begin{lemma}	\label{A2 lemme 3}
The following bounds holds uniformly in~$\vk$:
\begin{gather}
\label{A2 3a}
    a)\quad \Vert \, \op{B}_{\vk} \opU^{(l)}(s) \opQ \Vert \leq \Vert \, \op{B}_{\vk} \opU^{(l)}(s) \Vert \leq P(\sqrt{s}), \qquad l=1,2,\ldots,5	\\
\label{A2 3b}
    b)\quad \Vert \, \op{B}_{\vk} \opU^{({\rm D})}(s) \opQ \Vert_1 \leq P(\sqrt{s}) \e{-s E_1}	\\
\label{A2 3c}
    c)\quad \Vert \, \op{B}_{\vk} \opU^{(0)}(t)\opQ \op{B}_{-\vk} \opU^{({\rm D})}(s) \opQ\Vert_1 \leq P(\sqrt{s}) P(\sqrt{t}) \e{-s E_1}.
\end{gather}
\end{lemma}
%%%%%%%%%%%%%%%%%%%%%%%%%%%%%%%%%%%%%%%%%%%%
As opposed to the bounds of lemma~\ref{A2 lemme 2}, the above bounds involve the first excited state energy~$E_1$ of the hydrogen atom because of the operator~$\opQ$. Applying on~\eqref{A2 **} these bounds in the same way as we applied the bounds of lemma~\ref{A2 lemme 2} on~\eqref{A2 141}, we obtain the announced result \eqref{A2 **}$=o(\rhoat)$.

We show now that the terms in the class~II grow at most like $P(\sqrt{\be})\exp[2\be\mu]$ and hence are also $o(\rhoat)$. Contrary to the case considered in section~A.1, the terms in class~II of order less than~6 in~$\opV$ do not cancel out, because of the presence of the two operators~$\opQ$. We write therefore $\opQ = \id - \op{P}$ and look first at the terms without any~$\op{P}$. Just as in the contribution of~$R_1$, exact cancellations occur now at order 0 to~5 in~$\opV$. The terms of order 6 to 10 in~$\opV$ (and still without~$\op{P}$) are most easily majorized in the functional integral representation. Since $|\exp[-i\vk\cdot\frac{m\pr}{M}\ver](\exp[i\vk\cdot\ver]-1)/|\vk\Vert\leq C |\ver|$ uniformly in~$\vk$, these terms are dominated, in absolute values, by $C^2\exp[2\be\mu]$ times the expression~\eqref{A2 c3} which grows polynomially with~$\sqrt{\be}$. It remains thus only to consider the terms in the class~II with at least one ground state projector~$\op{P}$. Consider for example the term with $\opU^{(l)}(\be-\tau)\opP$ (the other terms can be treated similarly). Using the cyclicity of the trace, we can majorize this term by the matrix element
\eq{
    C \frac{\e{2\be\mu}}{(2\pi\lamat)^{3/2}} \int_0^\be\romd\tau\,
    |\bra{0} \op{P} \opB_{-\vk} \opU^{(m)}(\tau) \opB_{\vk} \opU^{(l)}(\be-\tau) \ket{0}|,
}
where $l,m=0,1,\ldots,5$. Thanks to the fast (exponential) decay of the ground state wavefunction, the operator $\opP\opB_{-\vk}$ (and its adjoint $\opB_{\vk}\opP$) are bounded uniformly in~$\vk$. If $l\neq0$, the matrix element involves therefore a product of bounded operators (see lemma~\ref{A2 lemme 3}.a and note that $\Vert\opU^{(m)}(s)\Vert\leq C\sqrt{s}$ as can be seen from the proof of lemma~\ref{A2 lemme 2}.a), and grows thus at most like a polynomial in~$\sqrt{\be}$. If $l=0$, one commutes $\opB_{\vk}$ with $\opU^{(0)}(\be-\tau)$ and rewrite the matrix element as
\eq{
    \bra{0} \op{P} \opB_{-\vk} \opU^{(m)}(\tau)  \opU^{(0)}(\be-\tau) \opB_{\vk} \opP \ket{0}
    +  \bra{0} \op{P} \opB_{-\vk} \opU^{(m)}(\tau) [\opB_{\vk},\opU^{(0)}(\be-\tau)] \ket{0}.
}
Since $[\opB_{\vk},\opU^{(0)}(\be-\tau)]$ is bounded (lemma~\ref{A2 lemme 8}.a), one can apply again the same argument. The terms in the class~II grow thus indeed at most like $P(\sqrt{\be})\exp[2\be\mu]$.

\subsection{Proof of the lemmas}

Here, we set units such that $e^2=1$ and $\hbar^2/m = 1$, so $\opV = -1/|\opq|$ and $\opU_0(s) = \exp[-s \opH_0]$. To prove lemmas~\ref{A2 lemme 2} and~\ref{A2 lemme 3}, we need a few basic results on norms involving the free propagators and the Coulomb operator:

\begin{lemma}	\label{A2 LEM UV}
The operator $\opU_0(s)\opV$ is bounded for $s>0$:
\eq{
    \Vert \opU_0(s)\opV \Vert \leq 2 (\frac{1}{\sqrt{s}}+2)
}
\end{lemma}

\begin{lemma}	\label{A2 LEM q_mu}
The following commutators are bounded:
\begin{gather}
\label{A2 [U_0,q]}
    a) \quad \Vert \, [\opU_0(s),\op{q}_\mu] \Vert \leq C \sqrt{s}	\\
\label{A2 [U_0,q]V}
    b) \quad \Vert \,[\opU_0(s),\op{q}_\mu] \opV\Vert \leq C (1 + \sqrt{s})	\\
\label{A2 [[U_0,q],q]V}
   c) \quad \Vert \, [[\opU_0(s),\op{q}_\mu],\op{q}_\nu] \opV \Vert \leq C \sqrt{s} (1 + \sqrt{s})
\end{gather}
\end{lemma}

\begin{lemma}	\label{A2 LEM UVUV}
The operator $\opU_0(s)\opV\opU_0(t)\opV$ belongs to the Hilbert-Schmidt class for $s,t>0$:
\eq{
    \Vert \opU_0(s)\opV\opU_0(t)\opV \Vert_2 \leq \frac{C}{\sqrt{s(s+t)}}
}
\end{lemma}

\begin{lemma}	\label{A2 LEM |q|}
The bounds of lemma \ref{A2 LEM q_mu} remain valid when $\op{q}_\mu$ (and $\op{q}_\nu$) is replaced by the operator $|\opq| = \sqrt{{\rm q}_1^2 + {\rm q}_2^2 + {\rm q}_3^2}$.
\end{lemma}

\begin{lemma}	\label{A2 lemme 8}
The following bounds hold uniformly in~$\vk$:
\begin{gather}
\label{A2 8a}
    a) \quad \Vert \, [\opB_{\vk},\opU_0(s)] \Vert \leq C \sqrt{s}	\\
 \label{A2 8b}
    b) \quad \Vert \, [\opB_{\vk},\opU_0(s) \opQ] \Vert \leq C (1 + \sqrt{s})	\\
\label{A2 8c}
    c) \quad \Vert \, [\opB_{\vk},\opU_0(s)] \opV\Vert \leq C (1 + \sqrt{s})	\\
\label{A2 8d}
    d) \quad \Vert \, [\opB_{\vk},[\opB_{-\vk},\opU_0(s)]] \opV \Vert \leq C \sqrt{s} (1 + \sqrt{s})
\end{gather}
\end{lemma}

The fact that the operator $\opU_0(s)\opV$ is bounded (lemma~\ref{A2 LEM UV}) can be traced back to the uncertainty principle of quantum mechanics. Indeed, if $\opV$ is very large, the wavefunction is close to the origin and hence well localized. By the uncertainty principle, the momentum must be large, so that $\opU_0(s)=\exp[-s \opp^2/2]$ is small. This makes the product $\opU_0(s)\opV$ bounded if $s>0$. Concerning lemma~\ref{A2 LEM UVUV}, recall that the Hilbert-Schmidt norm is defined as $\Vert\opA\Vert_2 = \sqrt{\Tr \opA\adj \opA}$. Naively speaking, lemma~\ref{A2 LEM UVUV} holds because the trace $\Tr \opA\adj \opA$ converges for the following two reasons: the product $\opU_0(s)\opV$ remains finite at short distances (lemma~\ref{A2 LEM UV}) and $\opV^4$ is integrable at large distances. In lemma~\ref{A2 LEM q_mu}.a, the presence of a commutator is crucial to ensure the finiteness of the operator norm (when $s=0$, this norm vanishes). In lemma~\ref{A2 LEM q_mu}.b, the commutator does not play such a crucial role, because ${\rm q}_\mu \opV$ is a bounded operator ($\Vert {\rm q}_\mu \opV \Vert \leq 1$). As already mentioned, the bounds of lemma~\ref{A2 lemme 8} are essentially the same as the bounds of lemma \ref{A2 LEM q_mu} and~\ref{A2 LEM |q|} because $\lim_{k\tend 0}\opB_{\vk} \sim \hat{\vk}\cdot\opq$. Lemmas \ref{A2 LEM UV}, \ref{A2 LEM q_mu} and \ref{A2 LEM UVUV} are established in~\cite{8bis}. We prove below the lemmas 2, 3 and 7, 8. 

\subsubsection*{Proof of lemma~\ref{A2 lemme 2}}

\noi a) We introduce~\eqref{A2 def U^(k)} in
\begin{multline}	\label{A2 191}
    \Vert\, |\opq| \opU^{(k)}(s) \Vert \leq \int_0^s\romd s_1 \ldots \int_0^{s_{k-1}} \romd s_k\,
    \Vert\, |\opq| \opU_0(s-s_1) \opV \Vert \\ \times
     \Vert\opU_0(s_1-s_2)\opV\Vert\cdot\ldots\cdot
    \Vert\opU_0(s_{k-1}-s_k)\opV\Vert \cdot \Vert\opU_0(s_k)\Vert.
\end{multline}
Notice that the operator $|\opq| \opU_0(s)\opV$ is bounded (this follows from lemma~\ref{A2 LEM |q|}.b):
\eq{	\label{A2 qU_0V}
     \Vert \, |\opq| \opU_0(s) \opV \Vert \leq C(1+\sqrt{s}).
}
Using lemma~\ref{A2 LEM UV} and $\Vert\opU_0(s_k)\Vert\leq 1$, \eqref{A2 191} involves integrals of the form
\eq{
    \int_0^{s_{k-1}} \romd s_k\, \frac{1}{\sqrt{s_{k-1}-s_k}} = 2 \sqrt{s_k} \, \leq \, 2\sqrt{s},
}
Hence only integrable singularities are present, and
$
    \Vert\, |\opq| \opU^{(k)}(s) \Vert \leq P_{2k}(\sqrt{s}),
$
where $P_{2k}(\sqrt{s})$ is a polynomial of order~$2k$ in~$\sqrt{s}$.\\

\noi b) From~\eqref{A2 def U^R}, one has
\begin{multline}
    \Vert \, |\opq| \opU\Reste(s) \Vert_1 \leq \int_0^s\romd s_1 \ldots \int_0^{s_5} \romd s_6\,
    \Vert \, |\opq| \opU_0(s-s_1) \opV \Vert \\ \times \Vert \opU_0(s_1-s_2)\opV\opU_0(s_2-s_3)\opV\opU_0(s_3-s_4)\opV\opU_0(s_4-s_5)\opV\Vert_1
\cdot \Vert \opU_0(s_5-s_6)\opV \Vert \cdot \Vert\opU(s_6) \Vert.
\end{multline}
The bound~\eqref{A2 2b} follows then from the use of $\Vert\opA\op{B} \Vert_1 \leq \Vert\opA\Vert_2\cdot\Vert\op{B}\Vert_2$ (the analogue of the Cauchy-Schwartz inequality for operators), lemma~\ref{A2 LEM UV}, \ref{A2 LEM UVUV}, \eqref{A2 qU_0V} and $\Vert\opU(s_6)\Vert\leq\exp[-s E_0]$. Note that the integrals are convergent. In particular
\eq{
    \int_0^{s}\romd s_1 \int_0^{s_1} \romd s_2 \int_0^{s_2} \romd s_3 \int_0^{s_3} \romd s_4\,
    \frac{1}{\sqrt{(s_1-s_2)(s-s_2)(s_3-s_4)(s_2-s_4)}} = 2 s^2.
}\\

\noi c) The trace norm in~\eqref{A2 2c} is majorized by
\begin{multline}
    \int_0^s\romd s_1 \ldots \int_0^{s_5} \romd s_6\,
    \Vert \, |\opq| \opU_0(t) |\opq|\opU_0(s-s_1) \opV \opU_0(s_1-s_2)\opV \Vert \\  \times
    \Vert \opU_0(s_2-s_3)\opV\opU_0(s_3-s_4)\opV\opU_0(s_4-s_5)\opV\opU_0(s_5-s_6)\opV\Vert_1 \cdot
 \Vert\opU(s_6) \Vert.
\end{multline}
Proceeding as in the proof of the point b), it remains only to show that
\eq{	\label{A2 fin}
    \int_0^s\romd s_1 \int_0^{s_1} \romd s_2\,
    \Vert \, |\opq| \opU_0(t) |\opq| \opU_0(s-s_1) \opV \opU_0(s_1-s_2)\opV \Vert
    \leq P(\sqrt{s}) P(\sqrt{t}).
}
To prove~\eqref{A2 fin}, we commute repeatedly the operators~$|\opq|$ to the right, making $|\opq|\opV=\id$ appear, or the bounded operator $|\opq|\opU_0(s) \opV$ (see~\eqref{A2 qU_0V}). We have thus, without specifying the time arguments,
\begin{multline}	\label{A2 184}
    |\opq|\opU_0|\opq|\opU_0\opV\opU_0\opV = 
    [|\opq|,\opU_0] |\opq| \opU_0 \opV \opU_0 \opV +
    \opU_0 |\opq| \opU_0  |\opq| \opV \opU_0 \opV  \\
    + \opU_0 [|\opq|,\opU_0]  |\opq|  \opV \opU_0 \opV +
    \opU_0 [|\opq|,[|\opq|,\opU_0]] \opV \opU_0 \opV.
\end{multline}
The bound~\eqref{A2 fin} follows then by applying the triangle inequality on~\eqref{A2 184} and using lemmas~\ref{A2 LEM UV} and~\ref{A2 LEM |q|}. \\

\subsubsection*{Proof of lemma~\ref{A2 lemme 3}}

The lemma~\ref{A2 lemme 3} can be proven exactly in the same way as lemma~\ref{A2 lemme 2}, if we use the bounds of lemma~\ref{A2 lemme 8} in place of the bounds of lemma~\ref{A2 LEM |q|}. The factor $\exp[-s E_1]$ comes from $\Vert\opU(s_6)\opQ\Vert\leq\exp[-s_6 E_1]\leq\exp[-s E_1]$.

\subsubsection*{Proof of lemma~\ref{A2 LEM |q|}}

\noi a) We use the upper bound $\Vert\opA\Vert \leq \max(S_{\vx},S_{\vect{y}})$, where $S_{\vx} = \sup_{\vx}\int\romd\vect{y} |\bra{\vx} \opA \ket{\vect{y}}|$, $S_{\vect{y}} = \sup_{\vect{y}}\int\romd\vect{x} |\bra{\vx} \opA \ket{\vect{y}}|$ and $\opA = [\opU_0(\tau),|\opq|]$:
\eq{
    S_{\vx} = S_{\vect{y}} = \sup_{\vect{y}}\int\romd\vx\, \frac{1}{(2\pi\tau)^{3/2}}
    	\e{-\frac{|\vx-\vect{y}|^2}{2\tau}} \absol{|\vx| - |\vect{y}| }.
}
We introduce the change of variables $\vx = \vect{u} + \vect{y}$, majorize $S_{\vect{y}}$ by inserting $\sup_{\vect{y}}$ under the integral sign and use $\sup_{\vect{y}} \Vert\vect{u}+\vect{y}| - |\vect{y}\Vert=|\vect{u}|$. We obtain thus
\eq{
    S_{\vect{y}} \leq \frac{1}{(2\pi\tau)^{3/2}} \int\romd\vect{u} \, \e{-\frac{\vect{u}^2}{2\tau}} |\vect{u}| \leq C \sqrt{\tau},
}
from which lemma \ref{A2 LEM |q|}.a follows.\\

\noi b) Since $|\opq|\opV = \id$, one has
\eq{
    \Vert [\opU_0(s),|\opq|] \opV \Vert = \Vert [\opU_0(s),\opq^2 \opV] \opV \Vert
	\leq \Vert \opU_0(s) \opq^2 \opV^2 \Vert + \Vert \opq^2 V \opU_0(s) \opV \Vert  
}
where we used the triangle inequality. The first norm is smaller or equal to one, and we majorize the second norm by
\eq{
    \Vert \opq^2 \opV \opU_0(s) \opV \Vert \leq \sum_{\mu=1}^3 \Vert q_\mu^2 \opV\opU_0(s)\opV \Vert
    \leq \sum_{\mu=1}^3 \Vert q_\mu\opV \Vert \cdot \Vert q_\mu \opU_0(s) \opV \Vert
    \leq C (1 + \sqrt{s}).
}
The last inequality follows from $\Vert q_\mu \opV\Vert\leq 1$ and lemma~\ref{A2 LEM q_mu}.b.\\

\noi c) We use $[[\opA,\op{B}],\op{B}] = [\op{B}^2,\opA] + 2[\opA,\op{B}]\op{B}$ and the triangle inequality to write
\eq{	\label{A2 qw}
    \Vert\, [[\opU_0(s),|\opq|],|\opq|]\opV \Vert \leq \Vert [\opq^2,\opU_0(s)] \opV \Vert + 2 \Vert [\opU_0(s),|\opq|] |\opq|\opV \Vert. 
}
From lemma~\ref{A2 LEM |q|}.a, the second norm in~\eqref{A2 qw} is dominated by $C\sqrt{s}$.
Since
\eq{
    [\opq^2,\opU_0]\opV = \sum_\mu(2 [{\rm q}_\mu,\opU_0] {\rm q}_\mu V
    	+ [{\rm q}_\mu,[{\rm q}_\mu,\opU_0]]),
}
we deduce from lemma \ref{A2 LEM q_mu}.a, \ref{A2 LEM q_mu}.c and $\Vert{\rm q}_\mu \opV\Vert\leq 1$ that the first norm in~\eqref{A2 qw} is bounded by $C\sqrt{s}(1+\sqrt{s})$, and hence lemma~\ref{A2 LEM |q|}.c is proven.

\subsubsection*{Proof of lemma~\ref{A2 lemme 8}}

We give the main steps in proving the points a) and b). The proof of the points c) and d) are left as an exercise to the reader.\\

\noi a) It is enough to show that
\eq{	\label{A2 carre}
    \Vert [\frac{\e{i\vk\cdot\opq}-1}{|\vk|},\opU_0(\tau)] \Vert \leq C \sqrt{s}, \qquad \forall\,\vk.
}
The operator $\exp[i\vk\cdot\opq]$ performs a translation in the space of momentum. Introducing the new notation $\opU^{(0)}_\tau(\vp)=\exp[-\tau \opp^2/2]$ for the free evolution operator, we have
\eq{
    \opU^{(0)}_\tau(\vp) \e{i\vk\cdot\opq} = \e{i\vk\cdot\opq} \opU^{(0)}_\tau(\vp + \vk).
}
The commutator in~\eqref{A2 carre} thus evaluates to $(\opU^{(0)}_\tau(\vp-\vk)-\opU^{(0)}_\tau(\vp)) \e{i\vk\cdot\opq}/|\vk|$. The bound~\eqref{A2 8a} follows then from $\Vert\exp[i\vk\cdot\opq]\Vert=1$ and
\eq{	\label{A2 abov}
    \frac{\Vert \opU^{(0)}_\tau(\vp-\vk) - \opU^{(0)}_\tau(\vp) \Vert}{|\vk|}
    \leq \frac{\sup_{\vp\in\IR^3}|\e{-\frac{\tau}{2}(\vp-\vk)^2} - \e{-\frac{\tau}{2}\vp^2}|}{|\vk|}
    \leq C \sqrt{\tau}.
}
In~\eqref{A2 abov}, the last inequality is uniform with respect to~$\vk$.\\

\noi b) Writing $\opQ = \id - \opP$, we have from the triangle inequality
\eq{
    \Vert [\opB_{\vk},\opU_0(s)\opQ] \Vert \leq \Vert [B_{\vk},\opU_0(s)] \Vert + \Vert [B_{\vk},\opU_0(s)\opP]  \Vert.
}
The first norm is bounded by $C \sqrt{s}$ according to lemma~\ref{A2 lemme 8}.a. The second norm is majorized by
$
    \Vert \opU_0(s)[B_{\vk},\opP] \Vert + \Vert [B_{\vk},\opU_0(s)] \opP \Vert.
$
We obtain the bound~\eqref{A2 8b} by using $\Vert\opP\Vert\leq1$, $\Vert\opU_0(s)\Vert\leq1$, lemma~\ref{A2 lemme 8}.a and $\Vert[\opB_{\vk},\opP]\Vert\leq C$.

\section{The dressed atomic contribution}	\label{S dat}
\newcommand{\chiTdat}{\TF{\chi}_{\rm d.at.}}

We calculate in this appendix the leading term in the atomic limit $\be\to\infty$ of the dressed atomic contribution~\eqref{A2 atomic ecrante}, by using the same method as in section~\ref{P atomic contribution}. Replacing $\phi$ by $V$ (using~\eqref{A2 132}), and omitting the factor $-\kappa^2/(k^2 + \kappa^2)$, we have to evaluate
\begin{multline}	\label{A2 atomic ecrante appendix}
    \chiTdat(\vk) \equiv -\frac{4\pi\be e^2}{k^2}
    \frac{2z\el}{(2\pi\lam\el^2)^{3/2}}\frac{2z\pr}{(2\pi\lam\pr^2)^{3/2}}
     \int\romd\ver \int\romD(\vxi\el)\int\romD(\vxi\pr)\int_0^1\romd s\int_0^1\romd t \\
	 \BB_{V,2}^{\TT\TT}(\ver,\chi\el,\chi\pr) 
    	 \Big(
    	\e{i\vk\cdot\lam\el\vxi\el(s)}\e{-i\vk\cdot\lam\el\vxi\el(t)} + \e{i\vk\cdot\lam\pr\vxi\pr(s)} \e{-i\vk\cdot\lam\pr\vxi\pr(t)} \\
    	-\e{-i\vk\cdot\ver} \e{i\vk\cdot\lam\pr\vxi\pr(s)} \e{-i\vk\cdot\lam\el\vxi\el(t)} 
    	-\e{i\vk\cdot\ver}  \e{i\vk\cdot\lam\el\vxi\el(s)} \e{-i\vk\cdot\lam\pr\vxi\pr(t)}
    	\Big)
    	 .
\end{multline}
We use the center of mass and relatives variables~\eqref{A2 Coord cm/rel}. The integrations over the center of mass factors, and is easily calculated using the covariance~\eqref{A2 cov}:
\eq{
    \frac{1}{(2\pi\lam\atH^2)^{3/2}} \int\romD(\vxi\atH) \e{i\vk\cdot\lam\atH[\vxi\atH(s)-\vxi\atH(t)]} = 
    \frac{1}{(2\pi\lam\atH^2)^{3/2}} \e{-\frac{1}{2} k^2 \lam\atH^2 |s-t|(1-|s-t|)} \equiv f_{|s-t|}(\vk).
}
We hence obtain the expression
\begin{multline}	\label{HYM}
   \chiTdat(\vk) = -\frac{4\pi\be e^2}{k^2} 4 \e{2\be\mu} \int_0^1\romd s \int_0^1 \romd t \, f_{|s-t|}(\vk) \frac{1}{(2\pi\lam^2)^{3/2}} \int\romd\ver\int\romD(\vxi) \\ \BB_{V,2}^{\TT\TT}(\ver,\lam\vxi) 
       \Big\{
   \e{i\vk\cdot\frac{m\pr}{M}\lam\vxi(s)} \e{-i\vk\cdot\frac{m\pr}{M}\lam\vxi(t)}
+ \e{-i\vk\cdot\frac{m\el}{M}\lam\vxi(s)} \e{i\vk\cdot\frac{m\el}{M}\lam\vxi(t)}\\
- \e{-i\vk\cdot\ver} \e{-i\vk\cdot\frac{m\el}{M}\lam\vxi(s)} \e{-i\vk\cdot\frac{m\pr}{M}\lam\vxi(t)}
- \e{i\vk\cdot\ver} \e{i\vk\cdot\frac{m\pr}{M}\lam\vxi(s) } \e{i\vk\cdot\frac{m\el}{M}\lam\vxi(t)}
    \Big\}.
\end{multline}
The integrand is a symmetrical function of $s$ and $t$, since $\BB_{V,2}^{\TT\TT}(\ver,\lam\vxi)=\BB_{V,2}^{\TT\TT}(-\ver,-\lam\vxi)$ and $\romD(\vxi)=\romD(-\vxi)$. The time integrations can hence be limited to the region $s>t$:
\eq{
   \int_0^1\romd s \int_0^1\romd t \to 2 \int_0^1\romd s\, \int_0^s \romd t.
}
It is convenient to factor the terms in the braces in~\eqref{HYM}, according to
\eq{
    \parenth{\e{i\vk\cdot\frac{m\pr}{M}[\ver+\lam\vxi(s)]}
    -\e{-i\vk\cdot\frac{m\el}{M}[\ver+\lam\vxi(s)]}}
    \parenth{\e{-i\vk\cdot\frac{m\pr}{M}[\ver+\lam\vxi(t)]}
    -\e{i\vk\cdot\frac{m\el}{M}[\ver+\lam\vxi(t)]}}.
}
In operatorial language, $\chiTdat(\vk)$ becomes
\begin{multline}	\label{A2 OKOK'}
    \TF{\chi}_{\text{d.at}}(\vk) = -\frac{4\pi e^2}{k^2} 4\e{2\be\mu} \frac{2}{\be}\int_0^\be\romd s \int_0^s \romd t\,   f_{|s-t|/\be}(\vk) \,
    \Tr \Big\{\e{-\be\opH}\opA\adj_{\rm int}(s)\opA_{\rm int}(t) \\
    -\e{-\be\opH_0}\opT \Big[ \opA\adj(s)\opA(t)
    \sum_{n=0}^5 \frac{1}{n!} \Big( \int_0^\be\romd u  \overline{\opV}(u) \Big)^n \Big] \Big\}.
\end{multline}
where $\opA_{\rm int}(s) = \opU(s)\opA\opU(-s)$ is the time evolved operator~$\opA$ (compare with~\eqref{A2 OKOK}). At low temperatures, the leading terms arise  from terms with ground state contributions from the evolution operator $\opU(\tau)=\e{-\tau \opH}$ ($\tau=t,s,\be$). Introducing the ground state projector $\opP=\ket{0}\bra{0}$ and $\opQ=\id-\opP$, the first term in the trace, i.e. $\e{-\be\opH}\opA\adj_{\rm int}(s)\opA_{\rm int}(t)$, can be written as (using the cyclicity of the trace)
\eq{
   \opU(\be-s+t) (\opP+\opQ) \opA \opU(s-t) (\opP + \opQ) \opA.
}
Retaining only terms with at least one ground state projector (as in~\eqref{A2 Ground State}), we find
\begin{multline}	\label{A2 OKOK P}
    \chiTdat^{(0)}(\vk) \equiv -\frac{4\pi e^2}{k^2} 4\e{2\be\mu} \frac{2}{\be}\int_0^\be\romd s \int_0^s \romd t\,   f_{|s-t|/\be}(\vk) \,
    \Tr \Big\{\opU(\be-s+t)\opP\opA\adj\opU(s-t)\opP\opA \\
    + \opU(\be-s+t)\opP\opA\adj\opU(s-t)\opQ\opA
    + \opU(\be-s+t)\opQ\opA\adj\opU(s-t)\opP\opA
     \Big\}.
\end{multline}
In terms of atomic eigenvalues and eigenstates, $\chiTdat^{(0)}(\vk)$ becomes
\begin{multline}	\label{ziz}
    \chiTdat^{(0)}(\vk) = -4\pi\rho\atH\ide \frac{e^2}{k^2} \frac{2}{\be} \int_0^\be\romd s \int_0^{s}\romd t\, \e{-\frac{1}{2}k^2\lam\atH^2\frac{s-t}{\be}(1-\frac{s-t}{\be})}
    \Big( |\opA_{00}(\vk)|^2\\
    +\sum_{m\geq 1}	\e{-(s-t)(E_m-E_0)} |A_{0m}(\vk)|^2
    +\sum_{m\geq 1}	\e{-(\be-s+t)(E_m-E_0)} |A_{0m}(\vk)|^2
    \Big)
\end{multline}
where we factored out the atomic density~\eqref{A2 def rho_at} and used $|\opA_{\n\n'}(\vk)|^2 = |\opA_{\n'\n}(\vk)|^2$. Let us evaluate~\eqref{ziz} for $k\ll\lam\atH^{-1}$. The first exponential can be approximated by one at lowest order and we find, after having performed the time integrations,
\eq{
   \chiTdat^{(0)}(\vk) \simeq -4\pi\rho\atH\ide \frac{2 e^2}{k^2} \sum_{\n=0}^{\infty} \frac{1-\e{-\be(E_{\n}-E_0)}}{E_{\n}-E_0} |\opA_{0\n}(\vk)|^2, \qquad k\lam\atH\ll 1.
}
We recover hence the polarisability $\alphaH(\vk,\beta)$ \eqref{A2 47}, so that eventually
\eq{	\label{B.resultat}
   \chiTdat^{(0)}(\vk) \simeq -4\pi\rho\atH\ide \alphaH,\qquad k\lam\atH\ll 1.
}

\end{document}